\documentclass[twocolumn,showpacs,showkeys,preprintnumbers,superscriptaddress,amsmath,floatfix,amssymb,secnumarabic,nofootinbib]{revtex4}
\usepackage[colorlinks=true]{hyperref}
\usepackage{graphicx}
\usepackage{braket}
\usepackage{amsmath}
\usepackage{epsfig}
\usepackage{float}
\usepackage{color}

\usepackage{bm}

\def\({\left(}
\def\){\right)}
\def\[{\left[}
\def\]{\right]}

\newcommand{\vev}[1]{ \langle \, #1 \, \rangle }

\newcommand{\Tr}{ {\rm Tr} \, }
\newcommand{\tr}{ {\rm Tr} \, }

\newcommand{\beq} {\begin{eqnarray}}
\newcommand{\eeq} {\end{eqnarray}}

\newcommand{\nn}{ \nonumber} 
\newcommand{\bpsi}{{\bar \psi}}
\newcommand{\bareta}{{\bar \eta}}

\newcommand{\comment}[1]{}

\begin{document}
\sloppy

\title{Collective charge excitations and the metal-insulator transition in the square lattice Hubbard-Coulomb model}

\author{Maksim Ulybyshev}
\email{maksim.ulybyshev@physik.uni-regensburg.de}
        \affiliation{Institut f\"ur Theoretische Physik, Universit\"at Regensburg, 93053 Regensburg, Germany}

\author{Christopher Winterowd}
\email{c.r.winterowd@kent.ac.uk}
\affiliation{University of Kent, School of Physical Sciences, Canterbury CT2 7NH, UK}

\author{Savvas Zafeiropoulos}
\email{savvas@jlab.org}
\affiliation{Jefferson Laboratory \\ 12000 Jefferson Avenue, Newport News, Virginia 23606, USA}
\affiliation{Department of Physics \\ College of William and Mary, Williamsburg, Virginia 23187-8795, USA}

\begin{abstract}
In this article, we discuss the non-trivial collective charge excitations (plasmons) of the extended square-lattice Hubbard model. Using a fully non-perturbative approach, we employ the hybrid Monte Carlo algorithm to simulate the system at half-filling. A modified Backus-Gilbert method is introduced to obtain the spectral functions via numerical analytic continuation.
We directly compute the single-particle density of states which demonstrates the formation of Hubbard bands in the strongly-correlated phase. The momentum-resolved charge susceptibility is also computed on the basis of the Euclidean charge density-density correlator.
In agreement with previous EDMFT studies, we find that at large strength of the electron-electron interaction, the plasmon dispersion develops two branches.
\end{abstract}

\maketitle

\section{Introduction}
The square-lattice Hubbard model has been the focus of intense theoretical research due to its simplicity and the fact that it demonstrates many of the phenomena that are associated with strongly-correlated electrons \cite{HubbardBook}. Furthermore, it is believed that in some sense, the physics of the high-temperature superconductors can be captured with the Hubbard model \cite{Hubbard_Cuprates_review}.

The Mott-Hubbard metal-insulator transition (MIT) \cite{MetalInsulatorReview}, the formation of a pseudogap \cite{Hubbard_Cuprates_review1, Pseudogap}, and the formation of Hubbard bands \cite{HubbardBands1,HubbardBands2} are all examples of strongly-correlated behavior that are expected to appear in the Hubbard model. In the case of the unfrustrated Hubbard model on the square-lattice, it is believed that strong, long-range antiferromagnetic fluctuations shift the MIT towards zero on-site coupling as the temperature is decreased towards zero \cite{FateMottHubbard}. Furthermore, these spin fluctuations evolve from being Slater-like at small to intermediate coupling to Heisenberg-like at large coupling.

More recent studies have extended the interactions in the Hubbard model beyond the on-site term and have included non-local correlations. These studies are closely related with the development of extended Dynamical Mean Field Theory (EDMFT) \cite{EDMFT1, EDMFT2} along with the EDMFT+$GW$ approach \cite{EDMFTGW}, where GW refers to the approximation of the self-energy by the first-order graph in which there appears one fermion line ($G$) and one screened interaction line ($W$) \cite{GWReview}. Among other phenomena, the inclusion of a long-range Coulomb interaction allows one to study collective charge excitations of the theory. These excitations, known as plasmons, can be accessed via the charge susceptibility $\chi_{\rho}({\bf q},\omega)$ which measures the system's response to a scalar potential $A_0({\bf q},\omega)$. Using the dual-boson approach, which goes beyond the EDMFT+GW approximation, it was found that at half-filling, the plasmons are characterized by a non-trivial dispersion relation \cite{vanLoon}. In this scenario, for a given strength of the Coulomb tail, at values of the on-site interaction $U$ close to the critical coupling for the metal-insulator transition, the plasmon dispersion separates into two branches as one approaches the edge of the Brillouin zone (BZ). It is argued that this feature can be viewed as a consequence of the formation of Hubbard bands.

Although this scenario seems plausible, it would be beneficial to have an independent, fully non-perturbative calculation. Due to the non-local nature of the Coulomb interaction, one can argue that existing methods may not in fact be accounting for all of the physics present in the system. A certain class of algorithms, going under the name of hybrid Monte Carlo (HMC) \cite{DuaneKogut}, are ideally suited for the calculations in strongly-correlated systems with non-local interactions \cite{ITEPRealistic}. Originally applied to the theory of the strong interactions, quantum chromodynamics (QCD), these methods have recently been applied successfully to certain condensed matter systems \cite{BuividovichUlybyshevReview,BuividovichPolikarpov,SmithVonSmekal,CCS0,CCS1, CCS2}. 
Such a fully non-perturbative calculation can be used not only as an independent check of the paper \cite{vanLoon} but also as a benchmark for further improvements of EDMFT methods.

In this paper, we perform calculations for the square lattice extended Hubbard model at half-filling using an interaction which includes an on-site term as well as a long-range ``Coulomb-tail" defined by the value of the nearest-neighbor interaction. Using a lattice Monte Carlo setup, we compute the single particle Green's function as well as the charge density-density correlator in Euclidean time. We then use these observables to obtain the density of states (DOS) and the charge susceptibility by directly performing the numerical analytic continuation (NAC). In doing so, we introduce a completely robust and generalized variant of the Backus-Gilbert (BG) method \cite{BackusGilbertOrig} for performing NAC. This scheme has recently been applied in studies of spectral functions of lattice quantum chromodynamics \cite{BrandtBG} and graphene \cite{UlybyshevConductivity}. Here we introduce an improved BG scheme based on the method of Tikhonov regularization \cite{Tikhonov}.

The remainder of the article is organized in the following way. In section \ref{sec:Setup}, we state our conventions and introduce the lattice setup used to perform the calculations. In section \ref{sec:Observables}, we outline the calculation of the fermion Green's function and charge density-density correlator. In section \ref{sec:AC}, we introduce the Green-Kubo (GK) relations and discuss, in general terms, the problem of obtaining real-frequency information from Euclidean correlators. From there, we describe our method for obtaining spectral functions and make comparisons with other closely related approaches. In section \ref{sec:Results}, we present our results for the charge susceptibility and the DOS and attempt to make contact with previous work \cite{vanLoon}. Finally, in section \ref{sec:Conclusion}, we draw conclusions and propose directions for further investigation.

\section{\label{sec:Setup}Lattice Setup}
\subsection{Extended Hubbard Hamiltonian}
We start by introducing the following tight-binding Hamiltonian on the square lattice
\beq \label{SqHubbard}
\hat{\mathcal{H}}_{\text{tb}} = - \kappa \sum_{\sigma} \sum_{\left \langle x, y \right \rangle} \left( \hat{c}^{\dagger}_{x,\sigma} \hat{c}_{y,\sigma} + \text{h.c.} \right),
\eeq
where $\kappa$ is the nearest-neighbor hopping parameter and the sum $\sum_{\left \langle x,y \right \rangle}$ runs over all pairs of nearest neighbors. The creation and annihilation operators satisfy the following anticommutation relations
\beq \label{CommRelations}
\left\{ \hat{c}_{x,\sigma}, \hat{c}^{\dagger}_{y,\sigma'} \right\} = \delta_{x,y} \delta_{\sigma, \sigma'},
\eeq
where $x,y$ refer to the lattice site and $\sigma, \sigma'$ refer to the electron's spin.
We now make the following canonical transformation on the creation and annihilation operators of the up- and down-spin electrons
\beq \label{ParticleHole}
\hat{c}_{x, \uparrow}, \hat{c}^{\dagger}_{x, \uparrow} &\to & \hat{a}_x, \hat{a}^{\dagger}_x, \\
\hat{c}_{x, \downarrow}, \hat{c}^{\dagger}_{x, \downarrow} &\to & \pm \hat{b}^{\dagger}_x, \pm \hat{b}_x,
\eeq
where the $\pm$ refers to whether the site $x$ is ``even" or ``odd". The lattice site $x$ is ``even'' if $(-1)^{x_1+x_2}=1$ and ``odd'' otherwise.
Thus, we can write (\ref{SqHubbard}), after a constant shift, as 
\beq
\hat{\mathcal{H}}_{\text{tb}} = - \kappa \sum_{\left \langle x, y \right \rangle} \left( \hat{a}^{\dagger}_x \hat{a}_y + \hat{b}^{\dagger}_x \hat{b}_y + \text{h.c.} \right).
\eeq
The Hilbert space of this tight-binding Hamiltonian can be constructed by first identifying the state satisfying $\hat{a}_x \ket{0} = \hat{b}_x \ket{0} = 0$ as the reference state. Thus, $\ket{0}$ corresponds to a state where each lattice site is occupied by one spin-down particle. The tight-binding Hamiltonian is written in momentum space as
\beq
\hat{\mathcal{H}}_{\text{tb}} = \sum_{{\bf k}} \epsilon_{{\bf k}} \left(  \hat{a}^{\dagger}_{{\bf k}} \hat{a}_{{\bf k}} + \hat{b}^{\dagger}_{{\bf k}} \hat{b}_{{\bf k}} \right),
\eeq
where $\epsilon_{{\bf k}} \equiv -2\kappa \sum_{i=1,2} \cos(k_ia)$, with $a$ being the lattice spacing. 

We now add two-body interactions between the electrons which are described by the term
\begin{equation}
  \hat{\mathcal{H}}_{\text{int}} =  \frac{1}{2}\sum_{x,y}  \hat{\rho}_x V_{x,y}\hat{\rho}_y,
\end{equation}
where $V_{x,y}$ is a positive-definite potential matrix and $\hat{\rho}_x$ is the electric charge operator at $x$ site which is defined as follows
\beq
\hat{\rho}_x \to \hat{a}^{\dagger}_x \hat{a}_x - \hat{b}^{\dagger}_x \hat{b}_x. 
\eeq
In our set up the matrix $V_{x,y}$ is defined completely by the on-site interaction (the Hubbard term) $U \equiv V_{0,0}$ and the nearest-neighbor interaction $V \equiv V_{(0,0), (0,1)} =  V_{(0,0), (1,0)}$.   
The latter coefficient characterizes the long-range $1/r$ Coulomb tail at any distance: $V_{x,y} = V/|\vec x-\vec y|,~x \neq y$, where the distance $|\vec x - \vec y|$ is dimensionless and evaluated in units of the lattice spacing. In order to obtain a positive-definite matrix $V_{x,y}$, these couplings must satisfy $U/V \gtrsim 1.5$. One also demands that the potential obeys periodic boundary conditions $V_{x+N_x,y} = V_{x,y+N_y} = V_{x,y}$, where $N_x$ and $N_y$ refer to the number of spatial lattice sites in the $x$ and $y$ directions. This slightly modifies the form of the potential relative to the infinite-volume form. Throughout this article, we will take $N_s \equiv N_x = N_y$.

\subsection{Path Integral Representation}
Following the approach of \cite{BuividovichPolikarpov,ITEPRealistic}, we start our construction of the path integral with the following Suzuki-Trotter decomposition of the partition function 
\beq \label{BoltzmanFactor} \nn
Z &\equiv& \Tr e^{-\beta\left( \hat{\mathcal{H}}_{\text{tb}} + \hat{\mathcal{H}}_{\text{int}} \right)}  = \Tr \left( e^{-\delta_{\tau} \left( \hat{\mathcal{H}}_{\text{tb}} + \hat{\mathcal{H}}_{\text{int}} \right)} \right)^{N_{\tau}} \\ &=& \Tr \left( e^{-\delta_{\tau} \hat{\mathcal{H}}_{tb}} e^{-\delta_{\tau} \hat{\mathcal{H}}_{\text{int}}} e^{-\delta_{\tau} \hat{\mathcal{H}}_{tb}} \dots \right) + O(\delta^2_{\tau}),
\eeq 
where $\beta \equiv 1/T$ and $\delta_{\tau} \equiv \beta/N_{\tau}$ defines the step in Euclidean time. To compute the trace, we insert resolutions of the identity using the Grassmann variable coherent state representation
\beq 
{\bm 1} = \int d\psi d\eta d \bpsi d \bareta e^{-\sum_x (\bpsi_x \psi_x + \bareta_x \eta_x )} \ket{\psi, \eta} \bra{\psi, \eta}, 
\eeq 
\vspace{-0.75cm}
\beq 
\ket{\psi, \eta} = e^{-\sum_x (\psi_x \hat{a}^{\dagger}_x +\eta_x \hat{b}^{\dagger}_x ) } \ket{0}.
\eeq 
Matrix elements of the form $\bra{\psi', \eta'} e^{-\delta_{\tau} \hat{\mathcal{H}}_{\text{tb}}} \ket{\psi, \eta}$ can be evaluated using the following identity 
\beq 
\bra{\psi'} e^{\sum_{x,y} \hat{a}^{\dagger}_x A_{x,y} \hat{a}_y} \ket{\psi} = \exp\left( \sum_{x,y} \bpsi'_x \left(e^A \right)_{x,y} \psi_y \right).
\eeq 
In order to apply this identity to the case of $\hat{\mathcal{H}}_{\text{int}}$, one must first perform the Hubbard-Stratonovich \cite{Hubbard,Stratonovich} transformation in order to obtain a bilinear in the exponent 
\beq  \nn
\exp\left( - \tfrac{\delta_{\tau}}{2} \sum_{x,y} \hat{\rho}_x V_{x,y} \hat{\rho}_y \right) \simeq
\eeq 
\vspace{-0.75cm}
\beq 
\int \mathcal{D}\phi \exp\left( - \tfrac{\delta_{\tau}}{2} \sum_{x,y}\phi_x V^{-1}_{x,y} \phi_y -i \delta_{\tau} \sum_x \phi_x \hat{\rho}_x  \right),
\eeq 
where $\phi_x$ is a real scalar field living on each site of the lattice. Putting all of this together, one finally arrives at the path integral representation of the partition function given by
\beq \label{PathIntegral1}
Z = \int \mathcal{D}\phi \mathcal{D}\psi \mathcal{D}\eta \mathcal{D}\bpsi \mathcal{D}\bareta e^{-\left(S_B[\phi] + \bpsi M[\phi] \psi + \bareta \bar{M}[\phi] \eta \right)},
\eeq 
where $S_B[\phi] = \tfrac{\delta_{\tau}}{2} \sum_{x,y,n}\phi_{x,n} V^{-1}_{x,y} \phi_{y,n}$ is the action of the Hubbard field and $n = 0,1,\dots,2N_{\tau}-1$ labels the factors of the identity that were inserted in (\ref{BoltzmanFactor}). We note that the Grassmann variables satisfy anti-periodic boundary conditions in Euclidean time. The fermionic operator $M$ is defined as follows
\beq \label{FermionAction} \nn
&& \sum_{x,y;\tau,\tau'} \bar{\psi}_{x,\tau} M_{x,y; \tau, \tau'} \psi_{y,\tau'}  = \\ && \nn \sum^{N_{\tau}-1}_{k=0} \biggl[ \sum_x \bar{\psi}_{x,2k} \left( \psi_{x,2k} - \psi_{x,2k+1} \right) \\ && \nn - \delta_{\tau} \kappa \sum_{\left \langle x, y \right \rangle} \left( \bar{\psi}_{x,2k} \psi_{y, 2k+1} + \bar{\psi}_{y,2k} \psi_{x,2k+1} \right) \\ && \nn+ \sum_x \bar{\psi}_{x,2k+1} \left( \psi_{x,2k+1} - e^{-i\delta \phi_{x,k}}\psi_{x,2k+2} \right) \biggr], 
\eeq 
where we have used the approximation $\exp\left( -\delta_{\tau} \hat{\mathcal{H}}_{\text{tb}} \right) \approx {\bm 1} - \delta_{\tau} \hat{\mathcal{H}}_{\text{tb}}$. The second fermionic term in the action is constructed using the relation $\bar{M} \equiv M^*$. It has been shown that the discretization errors present in the action (\ref{FermionAction}) are $O(\delta_{\tau})$ \cite{SmithVonSmekal}. 

The integration over the Grassmann variables in (\ref{PathIntegral1}) can be performed to obtain the following form of the partition function
\begin{equation} \label{PathIntegral2}
  Z = \int \mathcal{D}\phi  |\det M[\phi]|^2 e^{-S_B[\phi]},
\end{equation}
where we have used the identity 
\beq
\det M[\phi] \det \bar{M}[\phi] = |\det M[\phi]|^2,
\eeq
which follows from particle-hole symmetry. Immediately, one recognizes that the form of (\ref{PathIntegral2}) defines a positive-definite measure. Thus, one can immediately apply the HMC algorithm to study various equilibrium properties of the system at half-filling.

For our lattice Hamiltonian, the fermionic operator can have zero eigenvalues in the presence of a nonzero Hubbard field. In lattice Quantum Chromodynamics (LQCD), one must avoid these so-called ``exceptional configurations", and this is one of the reasons why a mass term for the fermions is introduced by hand \textit{e.g.} for studies regarding the properties of QCD in the deep chiral limit, such as the order of the chiral phase transition. As a consequence of this, one typically needs to extrapolate results to the chiral limit, $m\to 0$, which can be computationally expensive. In the present case, fermionic zero modes lie on an $(N-2)$-dimensional space where $N \equiv N^2_s \times N_{\tau}$ is the dimension of the space of Hubbard fields. This result can be seen by noting that the fermionic determinant $\det M(\phi)$ is a complex number. Thus, the two conditions for the appearance of a zero mode are $\mbox{Re} \det M(\phi) = \mbox{Im} \det M(\phi) = 0$. The fact that the fermionic determinant is a complex number is important here, since fermionic zero modes form an $(N-1)$ dimensional subspace in the case of a purely real fermionic determinant, where only one condition survives \cite{PhysRevB.90.035134}. 
As a result, for the complex fermionic determinant, these configurations are avoided in the molecular dynamics evolution and cannot divide the phase space into isolated regions. This is in direct contrast with LQCD where the phase space is divided into regions with different values for the topological charge \cite{DeTarDeGrand}. Thus, we do not need to introduce a mass term in our lattice action to obtain ergodic sampling of the phase space.

\section{\label{sec:Observables}Observables}
 Using the form of the partition function developed in the previous section, one can immediately write down an expression for the thermal expectation value of an operator $\mathcal{O}$
 \beq 
 \vev{ \mathcal{O}}_{\phi} = \frac{1}{Z} \int \mathcal{D}\phi |\det M[\phi]|^2\mathcal{O} e^{-S_B[\phi]}.
 \eeq 
 To access the single-particle DOS we calculate the spatial trace of the fermion Green's function 
 \beq \label{GreensFunction} \nn 
 G(\tau) &\equiv& -\sum_x \vev{ \hat{a}_{x}(\tau) \hat{a}^{\dagger}_x(0) }_{\phi} \\ &=& \sum_x \vev{ M^{-1}_{x,\tau;x,0} }_{\phi}, ~\tau=0,2,\dots,2N_{\tau},
 \eeq 
 where $\vev{}_{\phi}$ means the averaging over configurations of Hubbard field generated with the statistical weight (\ref{PathIntegral2}). In practice, one evaluates the trace on the right-hand side of (\ref{GreensFunction}) by the use of complex, Gaussian-distributed, stochastic vectors. It typically suffices to use $O(300)$ of these vectors on each configuration for lattice size $N_s=20$. 
 
 Since the single-particle DOS for half-filled system is symmetric with respect to zero, we will use the following symmetric Green-Kubo relation which connects the single-particle, momentum-averaged DOS $A(\omega) = \mbox{Im} G_R (\omega) / \pi$ to the Green's function in Euclidean time\cite{KadanoffMartin,MeyerTransport}
 \begin{equation} \label{GreenKuboDOS}
   G(\tau) = \int_0^\infty d\omega K(\tau,\omega) A(\omega),
 \end{equation}
 where the kernel for a correlator of the form $\vev{ \mathcal{O}(\tau)\mathcal{O}^{\dagger}(0)}$ is given by 
 \beq 
 K(\tau,\omega) \equiv \frac{\cosh\left[\omega(\tau-\beta/2)\right]}{\cosh(\omega \beta/2)}.
 \eeq 
  In the next section we will discuss in detail our method for inverting the relation in (\ref{GreenKuboDOS}) for $A(\omega)$. 
 
 To understand the collective charge excitations of the system, we calculate the response of the equilibrium system, to linear order, to an external potential $A^{(\text{ext})}_0({\bf r},t)$. The scalar potential couples linearly to the charge density $\hat{\rho}({\bf r})$. The deviation of the charge density from its equilibrium value due to this time-dependent perturbation is then expressed in momentum space as:
 \begin{equation}
    \vev{ \delta \hat{\rho}({\bf q},\omega)} = \chi_{\rho}({\bf q},\omega) A^{(\text{ext})}_0({\bf q},\omega),
 \end{equation}
 where we introduced the charge susceptibility  $\chi_{\rho}({\bf q},\omega)$. Translational invariance should be assumed to derive this relation. 

  From the charge density susceptibility, one can obtain the dielectric function
 \begin{equation}
   \frac{1}{\epsilon( {\bf q},\omega )} = 1 + V({\bf q}) \chi_{\rho}({\bf q},\omega ),
 \end{equation}
 where $V({\bf q})$ is the Fourier transform of the electron-electron interaction potential $V_{x,y}$. The peaks in ${\epsilon( {\bf q},\omega )}^{-1}$ give the dispersion relation for collective charge excitations (plasmons).
 
 The quantities above are all defined in real time (frequency). However, in our approach one computes the following Euclidean correlator
 \beq \nn 
   C({\bf q},\tau) &\equiv&  \vev{ \hat{{\rho}}_{{\bf q}}(\tau) \hat{{\rho}}_{{\bf q}}(0) } \\ &\equiv& \frac{1}{Z} \tr \left( e^{\tau \hat{\mathcal{H}}} \hat{{\rho}}_{{\bf q}} e^{-\tau \hat{\mathcal{H}}} \hat{{\rho}}_{{\bf q}} e^{-\beta \hat{\mathcal{H}}} \right),
 \eeq 
 where $\hat{\rho}_{{\bf q}} = \frac{1}{V} \sum_{{\bf x}} e^{-i {\bf q} \cdot {\bf x}} \hat{\rho}_{{\bf x}}$. By using the definition of $\hat{\rho}_{{\bf x}}$ and the properties of the coherent states, one can obtain an expression for the correlator in the path integral representation. The result is given by 
 \begin{equation}
    C({\bf q},\tau) = 2\sum_{{\bf x}} C({\bf x},\tau) \cos({\bf q} \cdot {\bf x}),
 \end{equation}
 where 
 \beq  \label{DensityDensityPositionSpace} \nn 
    &&C({\bf x} - {\bf x}',\tau) \equiv \vev{ \hat{\rho}_{{\bf x}}(\tau) \hat{\rho}_{{\bf x}'}(0) } = \\ \nn &&  - 2 \mbox{Re} \vev{ M^{-1}_{x,\tau;x',0} M^{-1}_{x',0;x,\tau} }_{\phi} + 2 \mbox{Re} \vev{ M^{-1}_{x,\tau;x,\tau} M^{-1}_{x',0;x',0} }_{\phi} \\ &&- 2 \mbox{Re} \vev{ M^{-1}_{x,\tau;x,\tau} \bar{M}^{-1}_{x',0;x',0} }_{\phi}.
 \eeq 
 In this expression we have performed an additional averaging over equivalent points in momentum space ($\pm \bf{q}$). The first term on the right-hand side of (\ref{DensityDensityPositionSpace}) is the connected piece while the other two terms constitute the disconnected part of the charge density-density correlator. Unlike the case of the current-current correlator \cite{UlybyshevConductivity}, both connected and disconnected parts are equally important, so the whole expression can not be calculated with a simple stochastic estimator. 
 
 The disconnected piece in (\ref{DensityDensityPositionSpace}) involves the correlation between two spatial traces evaluated on time slices separated by a distance $\tau$ in Euclidean time. It is thus necessary to perform $O(N), ~N \equiv N^2_s \times N_{\tau}$, inversions on each $\phi$ configuration as these pieces involve a fermion propagating from an arbitrary lattice point back to the same point. In LQCD, a variety of techniques have been employed to deal with the equivalent situation where quark disconnected loops are needed to accurately calculate an observable \cite{LiuStochastic, BaliStochastic, Gambhir:2016uwp}. 
  In this work, we have used a non-iterative solver based on the idea of Schur domain decomposition \cite{Schur_initial}. The solver is then applied to the point sources instead of the usual Gaussian-distributed stochastic ones. At the heart of the Schur complement method is the LU decomposition of dense matrix blocks contained inside the initial fermion operator matrix. The LU decomposition is made only once for a given $\phi$ configuration and is used repeatedly for all point sources. This allows us to work more efficiently in comparison with the commonly employed iterative solvers such as the conjugate gradient method.

 Just as in the case of the fermion Green's function, there exists a Green-Kubo relation which connects the Euclidean charge density-density correlator and its spectral function
 \begin{equation} \label{GreenKuboDensity}
   C({\bf q},\tau) = \frac{1}{\pi}\int d\omega K_\chi(\tau,\omega) \mbox{Im} \chi_{\rho}({\bf q},\omega),
 \end{equation}
 where the kernel for a correlator of the form $\vev{ \mathcal{O}(\tau)\mathcal{O}(0)}$ is given by
 \begin{equation} \label{DensityKernel}
   K_\chi(\tau,\omega) \equiv \frac{\cosh\left[\omega(\tau-\beta/2)\right]}{\sinh(\omega \beta/2)}.
 \end{equation}
 Thus, in full analogy with the paper \cite{vanLoon}, we can plot $\mbox{Im} {\epsilon( {\bf q},\omega ) }^{-1}$ in order to reveal the dispersion relation for the plasmons. 
In practice, it is more convenient to first solve the equation with the same kernel as for DOS
\begin{equation} \label{GreenKuboDensity1}
   C({\bf q},\tau) = \int d\omega K(\tau,\omega) \tilde\chi_{\rho}({\bf q},\omega),
 \end{equation}
and perform the rescaling for the spectral functions:
\begin{equation} \label{GreenKuboRescaling}
   \mbox{Im} \chi_{\rho}({\bf q},\omega) = \pi \tanh(\omega \beta /2)  \tilde\chi_{\rho}({\bf q},\omega),
 \end{equation}
Along with the case of DOS, (\ref{GreenKuboDensity}) will be the object of study in the coming sections.

 \section{\label{sec:AC}Analytic Continuation}
The central problem of this paper is obtaining the spectral functions for the fermion Green's function (DOS) and the charge density-density correlator in order to investigate the relationship between the formation of Hubbard bands and the nontrivial dispersion of the plasmons. However, directly inverting the relations in (\ref{GreenKuboDOS}) and (\ref{GreenKuboDensity}) constitutes an ill-posed problem. This is due to the fact that the kernel $K$ in (\ref{GreenKuboDOS}) has a very large condition number and thus even small changes in the Euclidean correlator $G(\tau)$  can lead to large changes in the spectral function $A(\omega)$ in frequency space. For this situation, a least-squares analysis is untenable.  

In the context of lattice QCD, several approaches to the solution of this problem have been used. Two prominent examples are the maximum entropy method (MEM) \cite{Hatsuda1,Hatsuda2,Swansea} and the Backus-Gilbert method \cite{BrandtBG}. MEM uses Bayes' theorem to regularize the inverse problem through the introduction of priors on the spectral function. In the end, one hopes to show that the resulting spectral function has little dependence on the form of the priors. It has been found that MEM can successfully identify sharp structures in frequency space, such as peaks, but can fail to identify other, more smooth features \cite{UlybyshevConductivity}.

 \begin{figure}
        \centering
         \includegraphics[scale=0.3, angle=-90]{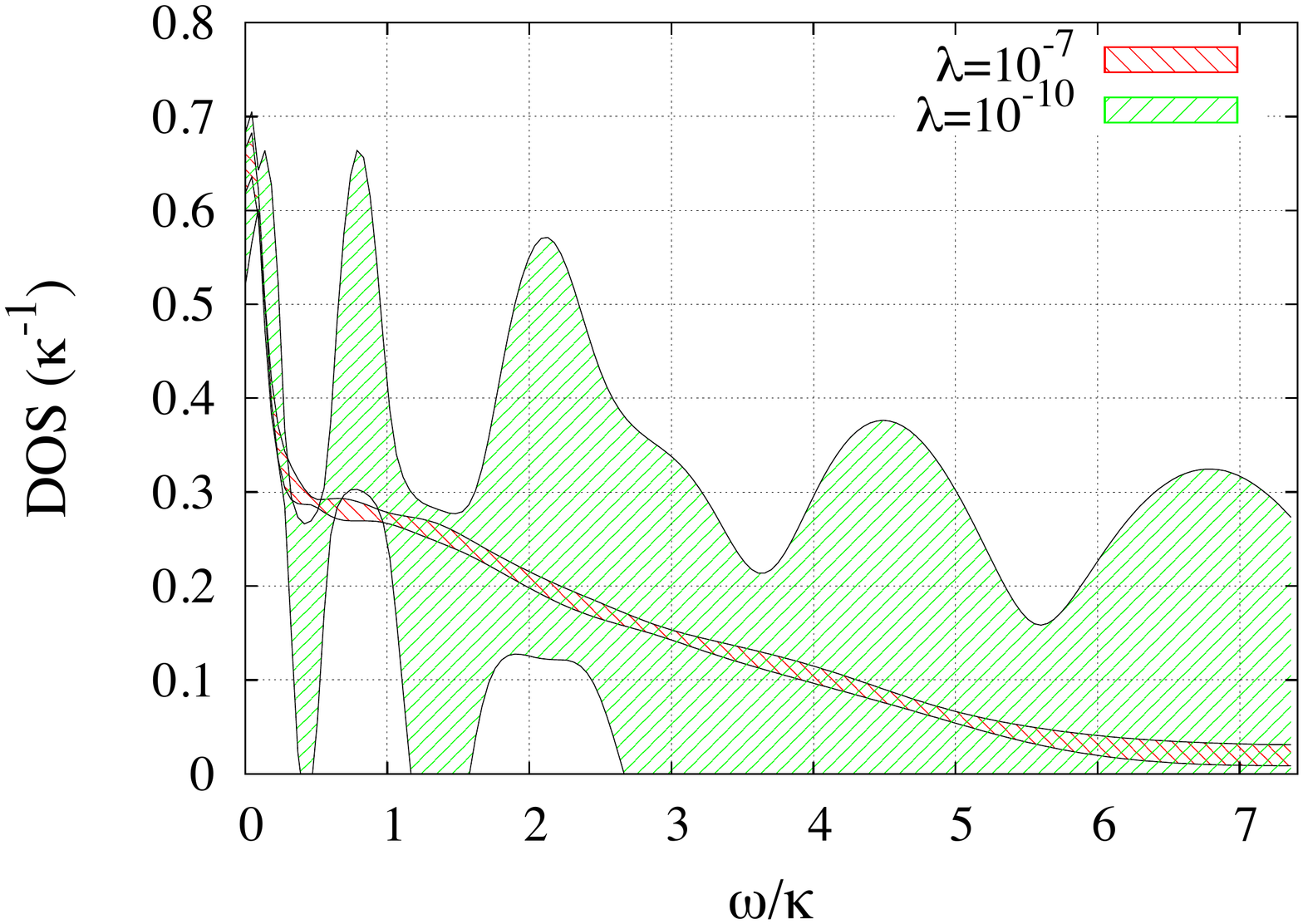}
        \caption{The effect of regularization on the spectral functions. We demonstrate the solution of (\ref{GreenKuboDOS}) for DOS using standard Tikhonov regularization. Here we plot the estimator for the spectral function obtained according to (\ref{ConvolutionDeltaFunction}). The frequency $\omega$ corresponds to the center of the resolution function and the filled area shows the statistical error. The example data is taken for the interaction strength $U/\kappa=1.66$, $V/\kappa = 0.62$ and temperature $T/\kappa = 0.046$.  The lattice with spatial size $N_s=20$ and $N_{\tau}=160$ Euclidean time slices is used. Resolution functions for $\lambda=10^{-7}$ can be found in Fig.~\ref{fig:DeltaFunctions}. }
        \label{fig:LambdaComparison}
\end{figure}
    
\begin{figure}
        \centering
          \includegraphics[scale=0.3, angle=270]{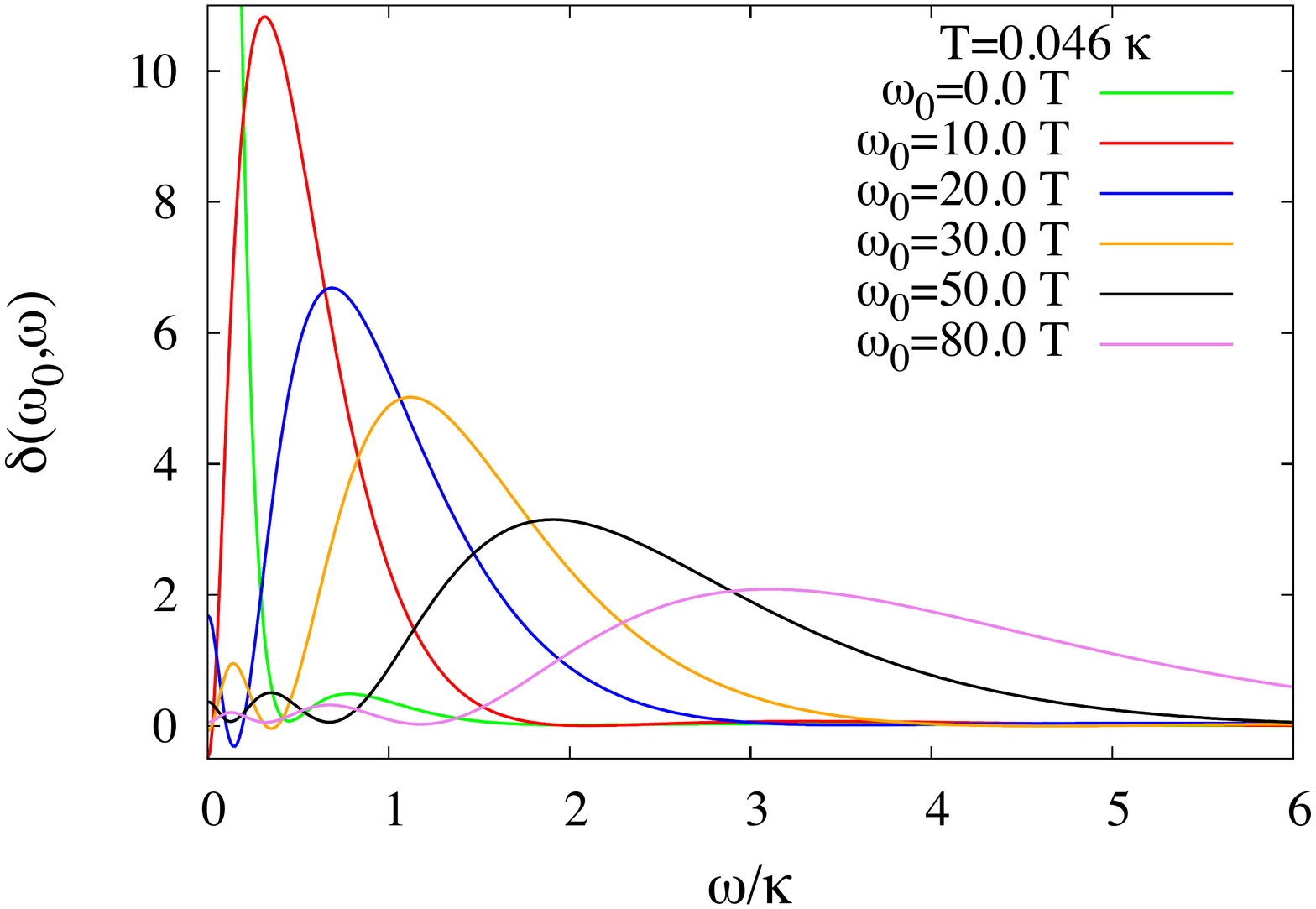}
        \caption{The resolution functions, $\delta(\omega_0,\omega)$, obtained with standard Tikhonov regularization (see (\ref{SVDStandardTikhonov})) during the solution of the Green-Kubo relation for DOS (\ref{GreenKuboDOS}). The parameter $\omega_0$ labels the center of the resolution function. The value of the regularization parameter is $\lambda=10^{-7}$. This set up will be used in all cases where we compute the DOS from the Monte Carlo data. }
        \label{fig:DeltaFunctions}
\end{figure}

 On the other hand, BG has been found to work well in characterizing the features of spectral functions in a variety of situations. The main advantage of this method is that one does not need to make assumptions about any particular feature of the spectral function. We will illustrate the use of this method via (\ref{GreenKuboDOS}). However, the discussion is not tied to any particular form of the kernel. The method starts by defining an estimator of the spectral function 
 \begin{equation} \label{ConvolutionDeltaFunction}
     \bar{A}(\omega_0) = \int^{\infty}_0 d\omega \delta(\omega_0,\omega) A(\omega).
 \end{equation}
 Thus, $\bar{A}(\omega_0)$ is the convolution of the exact spectral function $A(\omega)$ with the resolution function $\delta(\omega_0,\omega)$. One expresses the resolution function in the following basis
 \begin{equation}
   \delta(\omega_0,\omega) = \sum_{j} q_j(\omega_0) K(\tau_j, \omega), 
 \end{equation}
 where the coefficients $q_j(\omega_0)$ will be determined shortly. This definition of the resolution functions introduces the second important feature of the BG method, linearity. Thus, the error estimation is much simpler and it opens up the possibility for other improvements which will be discussed below in the text. Due to the linearity of the GK relations (see (\ref{GreenKuboDOS}) and (\ref{GreenKuboDensity})), one obtains
 \begin{equation} \label{Estimator}
   \bar{A}(\omega_0) = \sum_j q_j(\omega_0) G(\tau_j),
 \end{equation}
 where $G(\tau)$ is a generic correlator in Euclidean time (the momentum dependence was been suppressed). The resolution in frequency space is determined by the width of the resolution function around $\omega_0$
 \begin{equation} \label{ResolutionFunctionWidth}
   D \equiv \int^{\infty}_0 d\omega (\omega-\omega_0)^2 \delta^2(\omega_0,\omega),
 \end{equation}
 where $\int^{\infty}_0 d\omega \delta(\omega_0,\omega) = 1$. The coefficients in (\ref{Estimator}) are determined by minimizing the width, $\partial_{q_j} D = 0$, keeping the norm of the resolution function fixed. The result of this minimization yields
 \begin{equation}
   q_j(\omega_0) =\frac{W^{-1}(\omega_0)_{j,k}R_k}{R_n W^{-1}(\omega_0)_{n,m}R_m},
 \end{equation}
 where 
 \beq \label{W}
   W(\omega_0)_{j,k} &=& \int^{\infty}_0 d\omega (\omega-\omega_0)^2 K(\tau_j,\omega) K(\tau_k,\omega),\\ R_n &=& \int^{\infty}_0 d\omega K(\tau_n,\omega). 
 \eeq 
 The matrix $W$ is extremely ill-conditioned, with condition number $C(W) \equiv \frac{\lambda_{\text{max}}}{\lambda_{\text{min}}} \approx O(10^{20})$. It is thus imperative to regularize the method in order to obtain sensible results for a given set of data $G(\tau_i)$ and its associated error $\Delta G(\tau_i)$. 
 
\begin{figure}
        \centering
       \includegraphics[scale=0.3, angle=-90]{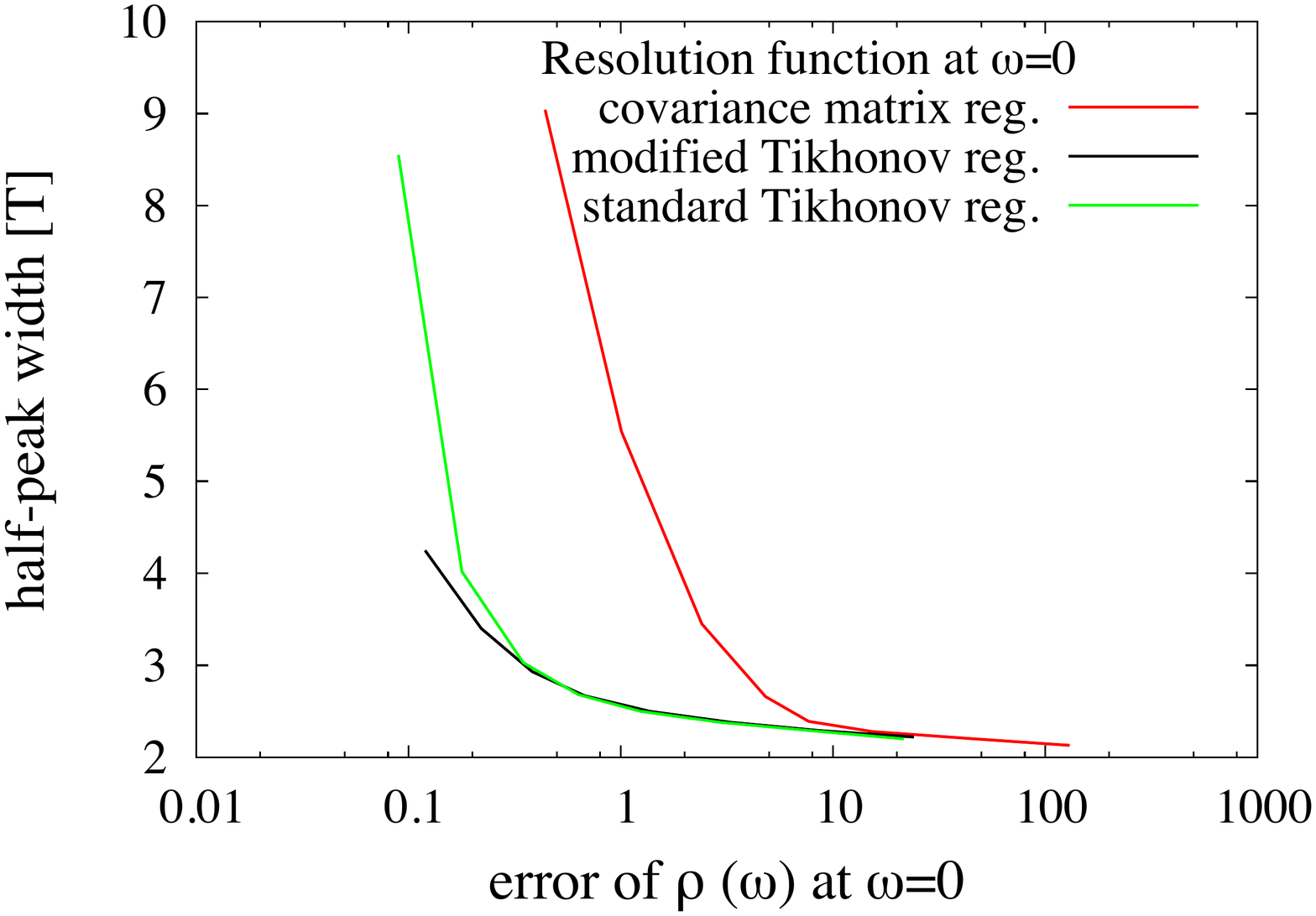}
        \caption{The half-peak width of the resolution function centered around $\omega_0=0$ versus the statistical error. The calculation is made with different regularization algorithms for the correlator $G(\tau)$ calculated on the lattice with spatial size $N_s=20$ and $N_{\tau}=160$ for the interaction strength $U/\kappa=1.66$, $V/\kappa = 0.62$ and temperature $T/\kappa = 0.046$. The statistical error is controlled by the regularization parameter $\lambda$. The resolution functions are computed in the process of inverting the Green-Kubo relation for DOS (\ref{GreenKuboDOS}). The width is given in units of temperature and one can see that as $\lambda$ vanishes each curve converges to $\sim 2$. This is as expected as the resolution in frequency space is ultimately limited by the temperature.}
        \label{fig:ResWidthComparison}
\end{figure}
   
 Previous studies employing the BG method have used the so-called ``covariance" regularization \cite{BrandtBG,UlybyshevConductivity}. In this approach, the following modification is made in (\ref{W})
 \begin{equation}
  W(\omega_0)_{j,k} \to (1-\lambda) W(\omega_0)_{j,k} + \lambda C_{j,k}, 
 \end{equation}
 where $\lambda$ is a small regularization parameter and $C_{j,k}$ is the covariance matrix of the Euclidean correlator $G$. The hope is that this replacement helps improve the condition number of the matrix $W$ while still maintaining a sufficiently small width of the resolution functions in frequency space.  
 
 Although covariance regularization performs well, one might wonder as to the merits of other commonly used regularization methods for ill-posed problems. Furthermore, in numerical studies where a covariance matrix cannot be constructed (i.e.~when the Euclidean correlator data are obtained using a non-stochastic procedure), covariance regularization cannot be applied. The regularization method that we propose, the so-called Tikhonov regularization \cite{Tikhonov}, is a widely used approach to ill-posed problems of the form $Ax=b$. In this method, one seeks a solution to the modified least-squares function
 \begin{equation}
   \text{min} \left( \Vert A x - b \Vert^2_2 + \Vert \Gamma x \Vert^2_2 \right),
   \label{Tikh_def}
 \end{equation}
 where $\Gamma$ is an appropriately chosen matrix.   The effect of various types of Tikhonov regularization on the matrix $W$ can be most easily seen by employing the singular value decomposition (SVD). In this procedure 
 \begin{equation}
   W = U \Sigma V^{\top}, ~UU^{\top} = VV^{\top} = {\bm 1},
 \end{equation}
 where $\Sigma = \text{diag}(\sigma_1,\sigma_2,\dots,\sigma_N),~\sigma_1 \geq \sigma_2 \geq \cdots \geq \sigma_N$. The inverse is thus easily expressed as 
 \begin{equation}
  W^{-1} = V D U^{\top},~D = \text{diag}( \sigma^{-1}_1,\sigma^{-1}_2,\dots,\sigma^{-1}_N ). 
 \end{equation}
In the standard Tikhonov regularization, one modifies the matrix D in the following way 
 \begin{equation} \label{SVDStandardTikhonov}
   D_{i,j} \to \tilde{D}_{i,j} = \delta_{ij}\frac{\sigma_i}{\sigma^2_i + \lambda^2},
 \end{equation}
  where $\lambda$ is again the regularization parameter. One can see that the singular values which satisfy $\lambda \gg \sigma_i$ are smoothly cut off. This procedure corresponds to $\Gamma = \lambda {\bm 1}$ in (\ref{Tikh_def}). One thus pays a price for solutions that are not ``smooth". In general, for small $\lambda$, the solutions fit the data well but are oscillatory, while at large $\lambda$, the solutions are smooth but do not fit the data as well.
 We have also tested an alternative method which regulates the small singular values of $W$ in a smoother fashion
  \begin{equation} \label{SVDModifiedTikhonov}
  \tilde{D}_{i,j} = \delta_{ij}\frac{1}{\sigma_i + \lambda}.
 \end{equation}
 For this choice, which we refer to as ``modified Tikhonov", we give preference to spectral functions which give smooth reconstructed Green's functions in Euclidean time. This method corresponds to   $\Gamma = \lambda A$ ($A \to K$ in our case) in (\ref{Tikh_def}). 
 
The effect of regularization on the reconstructed spectral function is shown in Fig.~\ref{fig:LambdaComparison} for the case of DOS. The corresponding resolution functions obtained with ordinary Tikhonov regularization with $\lambda=10^{-7}$ are plotted in Fig.~\ref{fig:DeltaFunctions}. One can clearly see that varying the regularization parameter $\lambda$ has a significant effect on the resulting spectral function. This set up will be used later for further calculations of DOS. 
 
\begin{figure}
        \centering
        \includegraphics[scale=0.3, angle=270]{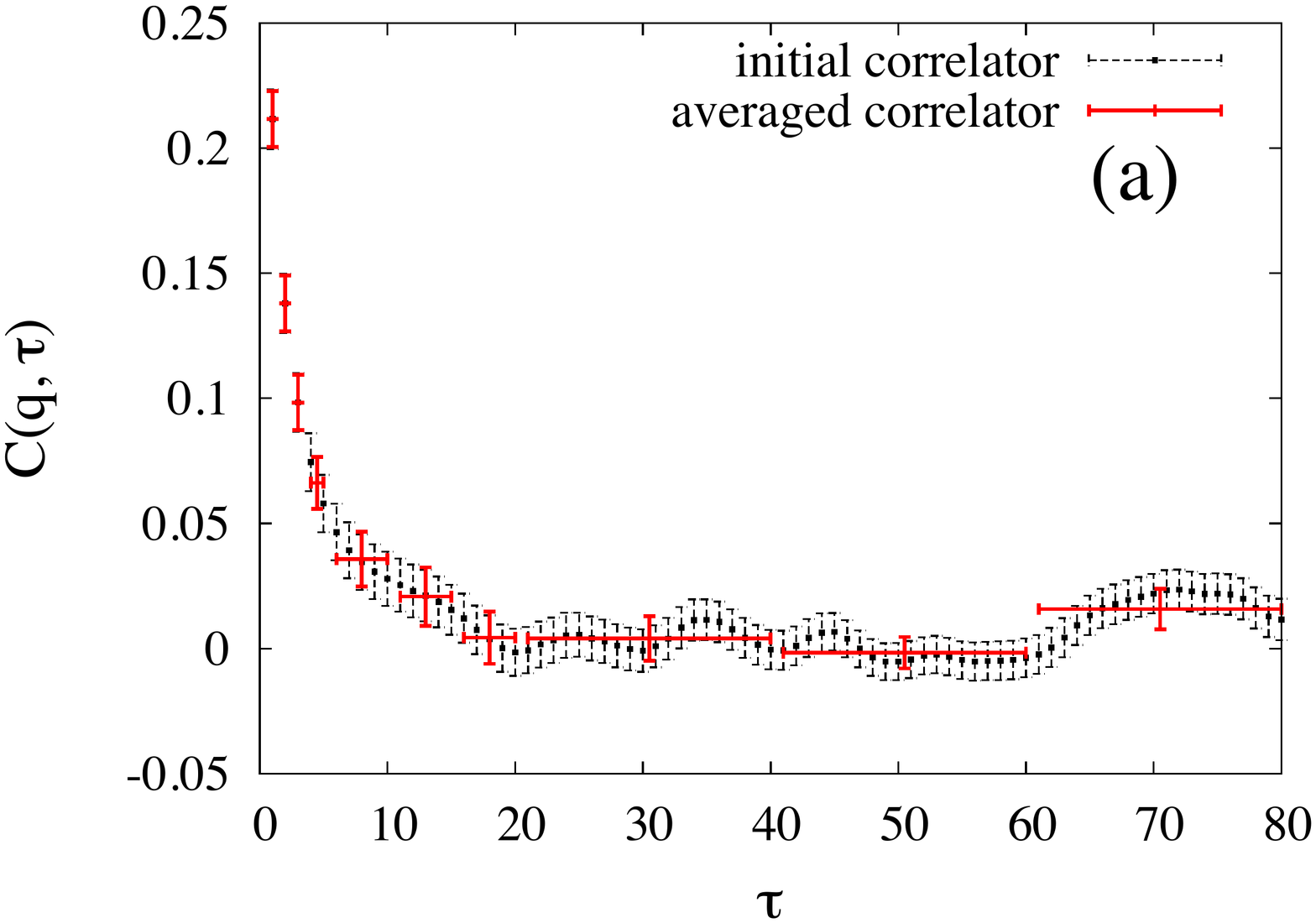}
        \includegraphics[scale=0.3, angle=270]{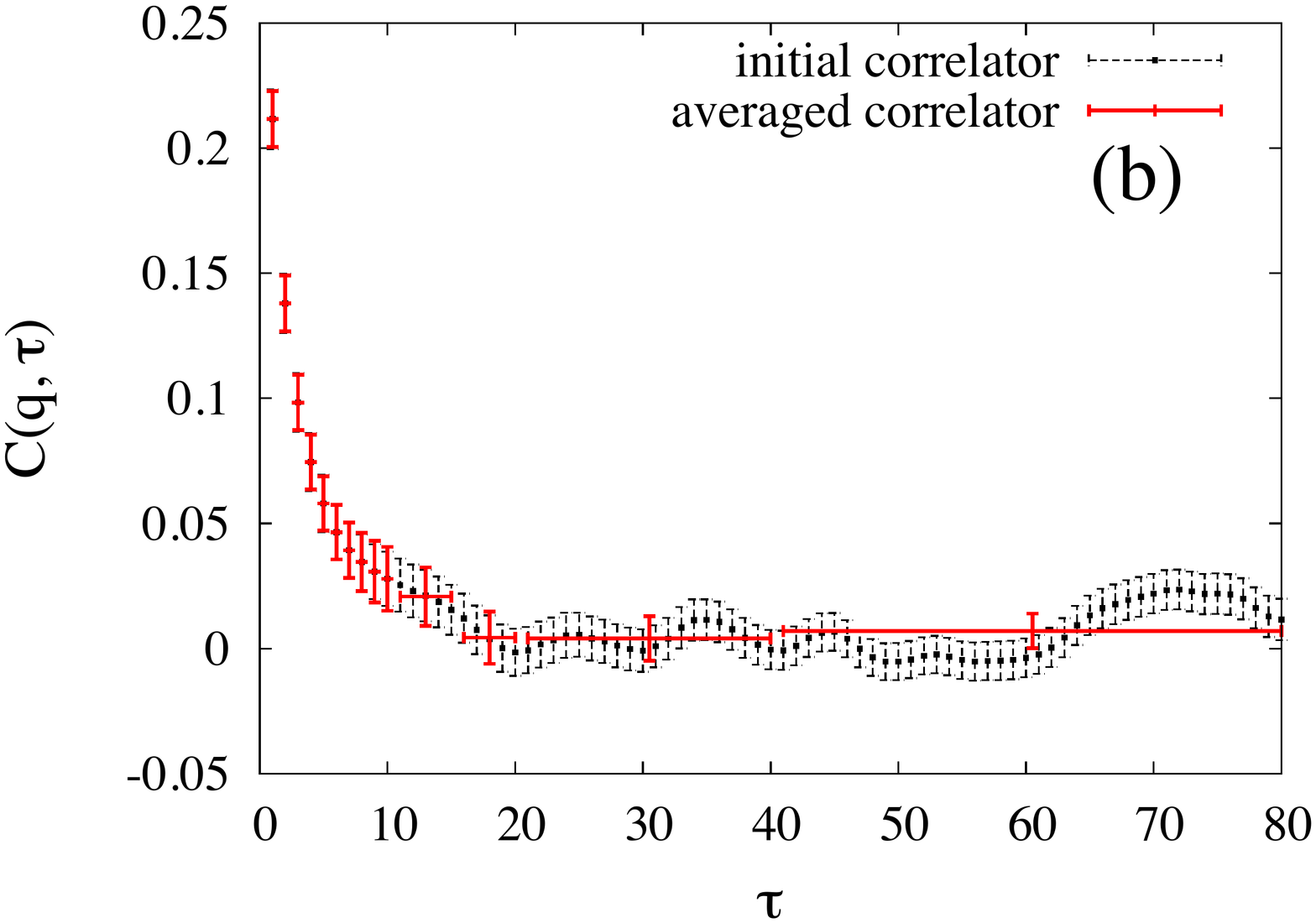}
        \caption{Two variants of correlator averaging. The charge density-density correlator $C({\bf q},\tau)$ is calculated for $U/\kappa=3.33$, $V/\kappa=1.26$ and $T/\kappa=0.046$ using a lattice with spatial size $N_s=20$ and $N_\tau=160$ Euclidean time slices. The momentum $q$ belongs to the $X-M$ line in the Brillouin zone, which is the most physically interesting case (see below in section \ref{sec:Results}).}
        \label{fig:corr1}
\end{figure}
\begin{figure}
        \centering
         \includegraphics[scale=0.3, angle=270]{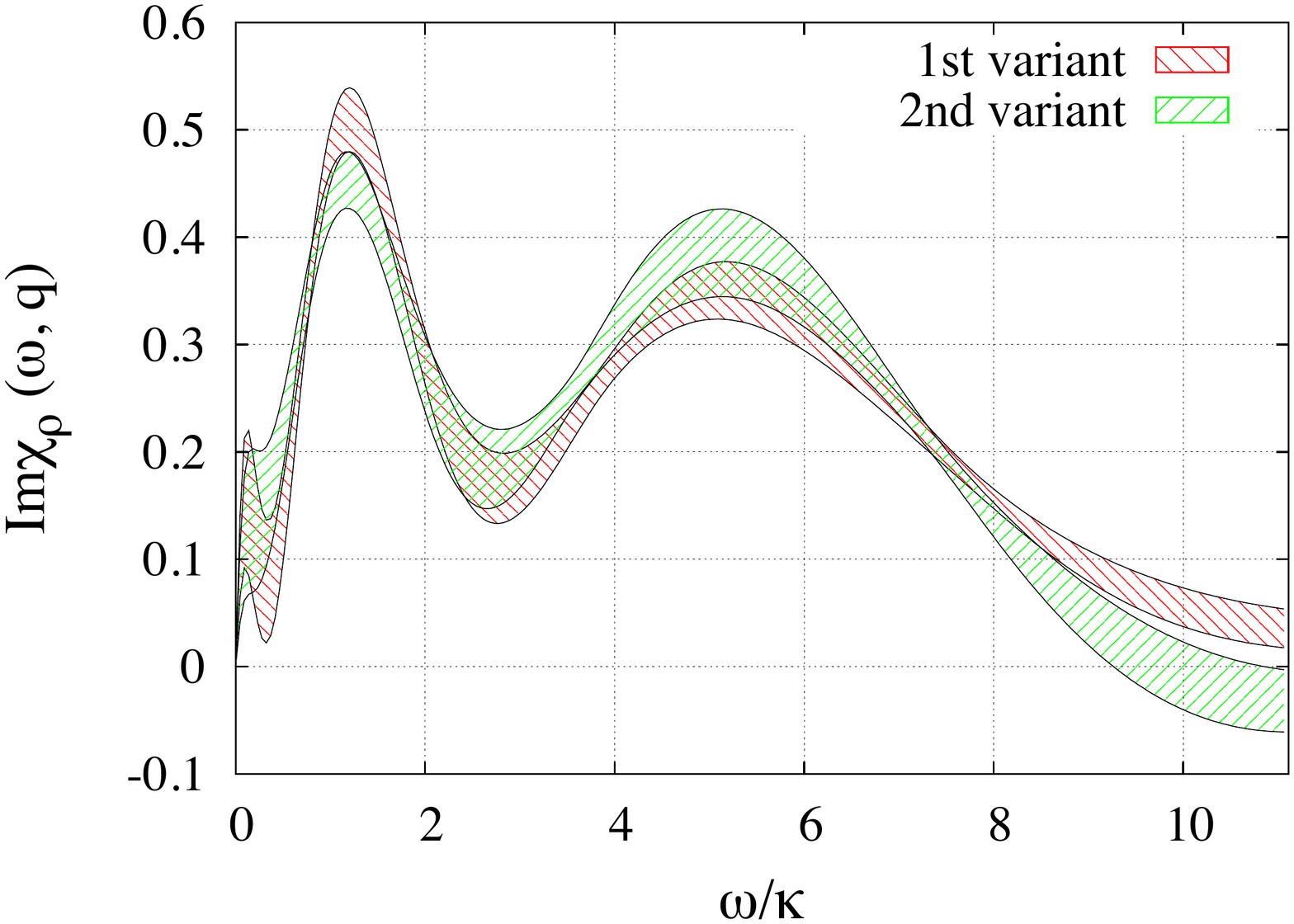}
              \caption{ Spectral functions $\mbox{Im} \chi_\rho (\omega,q)$ computed using the relations (\ref{GreenKuboDensity1}) and (\ref{GreenKuboRescaling}) for two different variants of the averaged correlator shown in Fig.~\ref{fig:corr1}. In both cases ordinary Tikhonov regularization is used with $\lambda=5.0 \times 10^{-6}$.
             We plot the estimator for the spectral function obtained according to (\ref{ConvolutionDeltaFunction}). The frequency $\omega$ corresponds to the center of the resolution function and the filled area shows the statistical error.}
        \label{fig:compare_average}
\end{figure}
To gain an accurate estimate of the statistical error in our spectral functions, we perform a data binning as follows. Taking our original ensemble of $N_{\text{conf}}$ Hubbard field configurations generated according to the weight (\ref{PathIntegral2}), we construct $\tilde{N}$ blocks of $N_{\text{conf}}/\tilde{N}$ configurations. We are then left with several subsets of Euclidean time correlators
\beq \nn 
&& \{ G^{(i)}(\tau_j), i=1,\dots,N_{\text{conf}} \} \to \\ \nn  &&\{ G^{(i)}(\tau_j), i=1,\dots,N_{\text{conf}}/\tilde{N} \}, \dots , \\ && \{ G^{(i)}(\tau_j), i=(\tilde{N}-1)N_{\text{conf}}/\tilde{N}+1,\dots,N_{\text{conf}} \}.
\eeq 
The number of blocks, $\tilde{N}$, is chosen by examining the autocorrelation of the Euclidean correlator between different Hubbard field configurations and enforcing the condition $N_{\text{conf}}/\tilde{N} \gg l_{\text{max}}$, where $l_{\text{max}}$ is the maximum autocorrelation length pertaining to  $G(\tau_i), i=0,1,\dots,N_{\tau}-1$. This condition ensures that the size of each block is such that it contains numerous statistically-independent configurations and each block is statistically independent of all other blocks. Using these blocks of correlators, we construct $\tilde{N}$ spectral functions $\bar{A}_i$ and calculate an average spectral function 
\beq
\bar{A}_{\text{avg}} \equiv \frac{1}{\tilde{N}} \sum^{\tilde{N}}_{i=1} \bar{A}_i, 
\eeq 
and its associated error $\sigma(\bar{A})$ for each frequency. We have found that this procedure yields a much better estimate of the statistical error than simply propagating the error in the Euclidean correlator to the spectral function through the relation in (\ref{Estimator}). 
Fig.~\ref{fig:LambdaComparison} demonstrates typical behaviour of the statistical errors when we switch on the regularization. If the regularization is not sufficient, the spectral function has huge statistical errors. Once we increase $\lambda$, the errors are suppressed and the spectral function converges to some stable average value. As mentioned previously (see Fig.~\ref{fig:ResWidthComparison}), the price for this ``smoothing'' is the enlargement of the width of all resolution functions which may imply the loss of some information.

We now compare all three types of regularization using the width of the resolution functions as a metric. The study is summarized in Fig.~\ref{fig:ResWidthComparison},  where we plot the width at half maximum of the resolution function with center at $\omega_0=0$ versus the statistical error of the reconstructed spectral function (again at $\omega_0=0$) for all regularization methods. The data for DOS were used as a test case. 
As one can see, both types of Tikhonov regularization work better in the sense that they provide better resolution in frequency being equally efficient in suppressing the statistical error. Or, in other words, for Tikhonov regularization the statistical error is smaller for the same resolution in frequencies. On can see also, that 
in practice the difference between the standard (\ref{SVDStandardTikhonov}) and modified Tikhonov (\ref{SVDModifiedTikhonov}) regularizations is negligible. As a consequence, we will use standard Tikhonov approach in all further calculations.  
 
The algorithm for choosing the optimal value of $\lambda$ is based on the ``global relative error" for the spectral function
\beq \label{GlobalRelativeError}
\mathcal{G} \equiv \frac{1}{N_0} \sum_{\omega_0} \frac{\sigma\left( \bar{A}(\omega_0) \right)}{\bar{A}_{\text{avg}}(\omega_0)}, 
\eeq 
where the sum in the above expression runs over the centers of the resolution functions and $N_0$ is the  number of resolution functions with different centers $\omega_0$. Our basic criteria for the choice of $\lambda$ is that the ``global relative error" should be within the interval  $5-10\%$. We thus sufficiently suppress the statistical error while still maintaining good resolution in frequency space. 

Using the quantity in (\ref{GlobalRelativeError}) as a measure, we start from small $\lambda$ and increase it until we have obtained the desired statistical error. Typically, we have taken $\lambda=10^{-7}-10^{-5}$ in obtaining the results in this paper. 

 \begin{figure}
        \centering
        \includegraphics[scale=0.28, angle=-90]{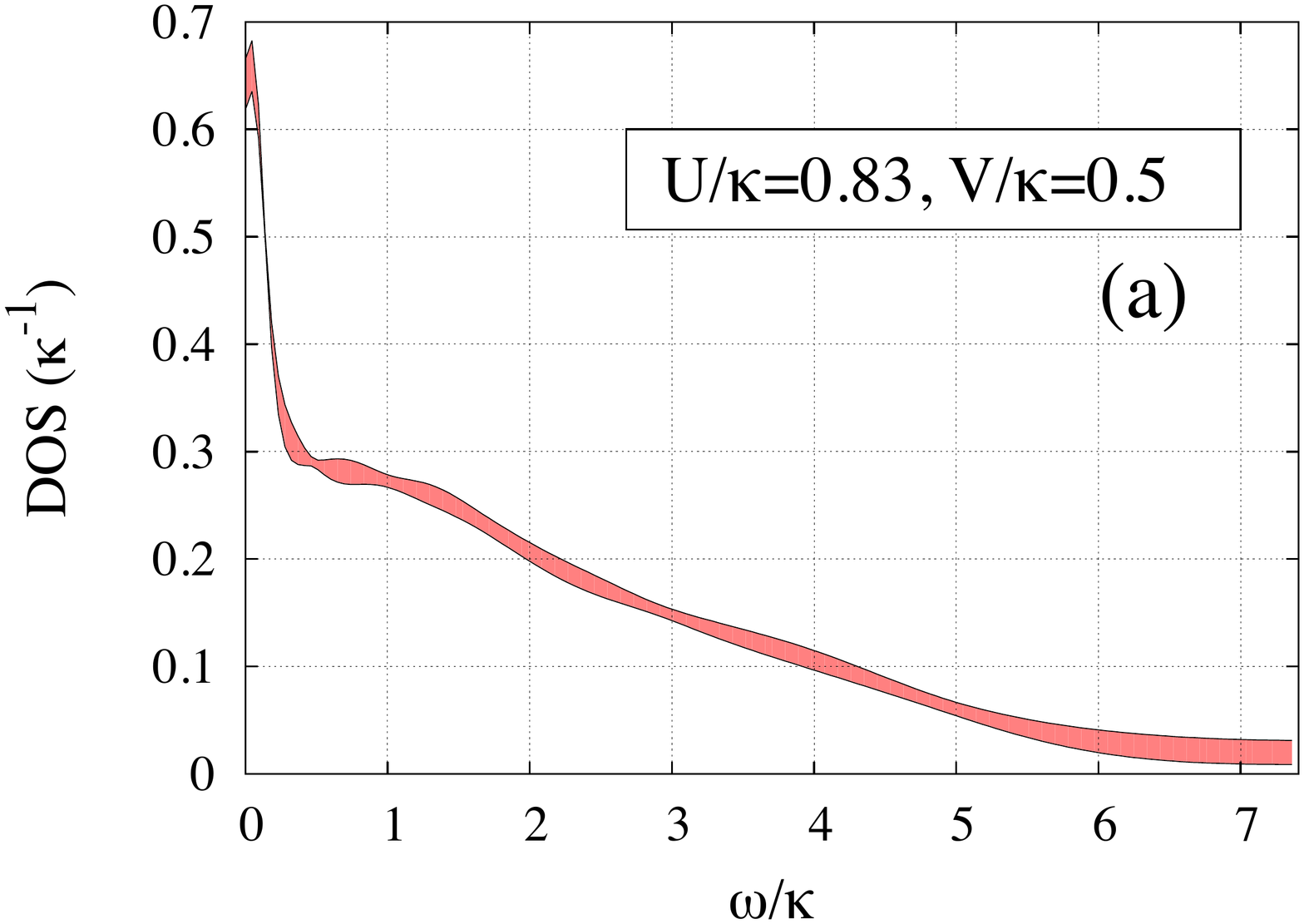}
        \includegraphics[scale=0.28, angle=-90]{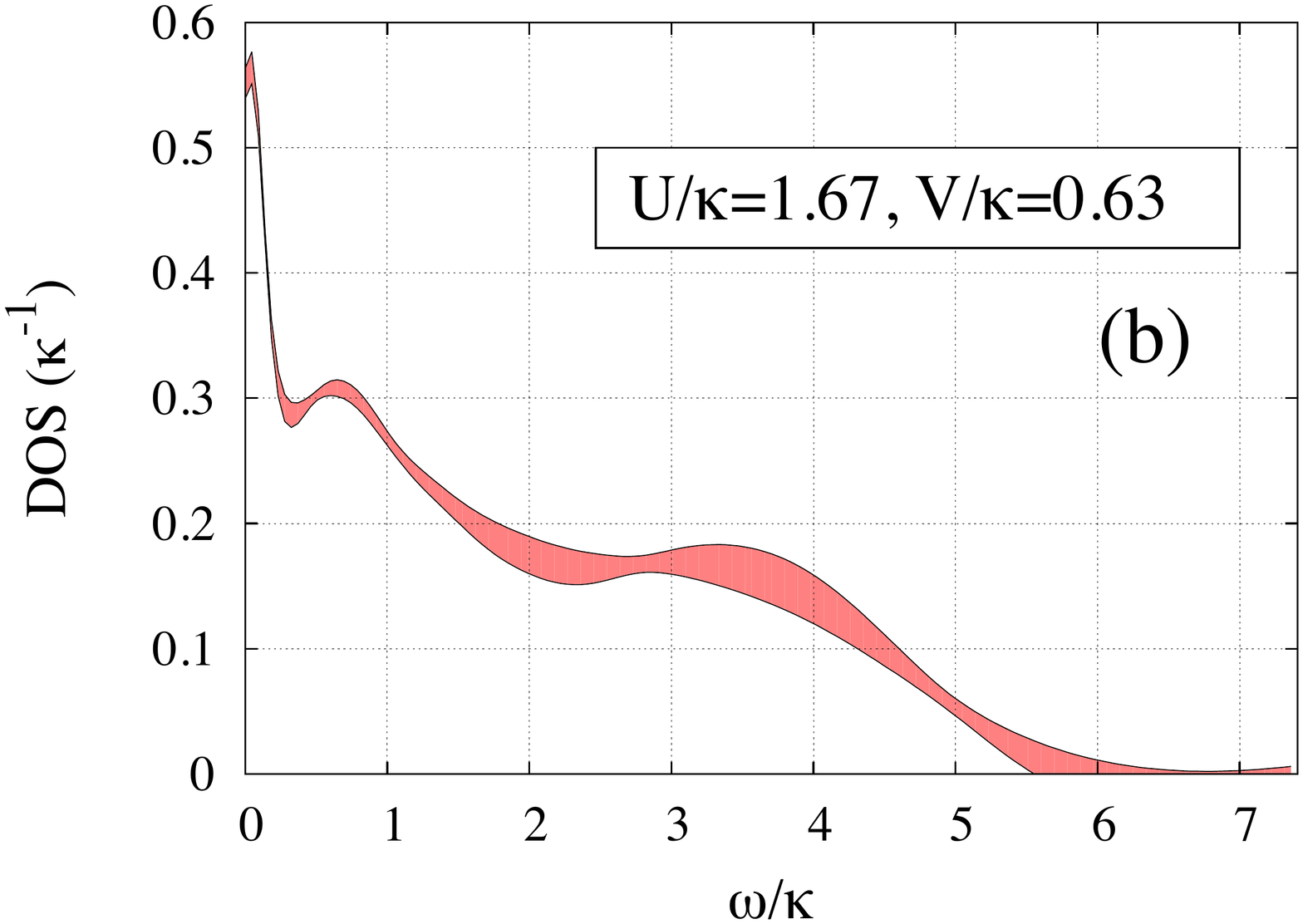}
        \includegraphics[scale=0.28, angle=-90]{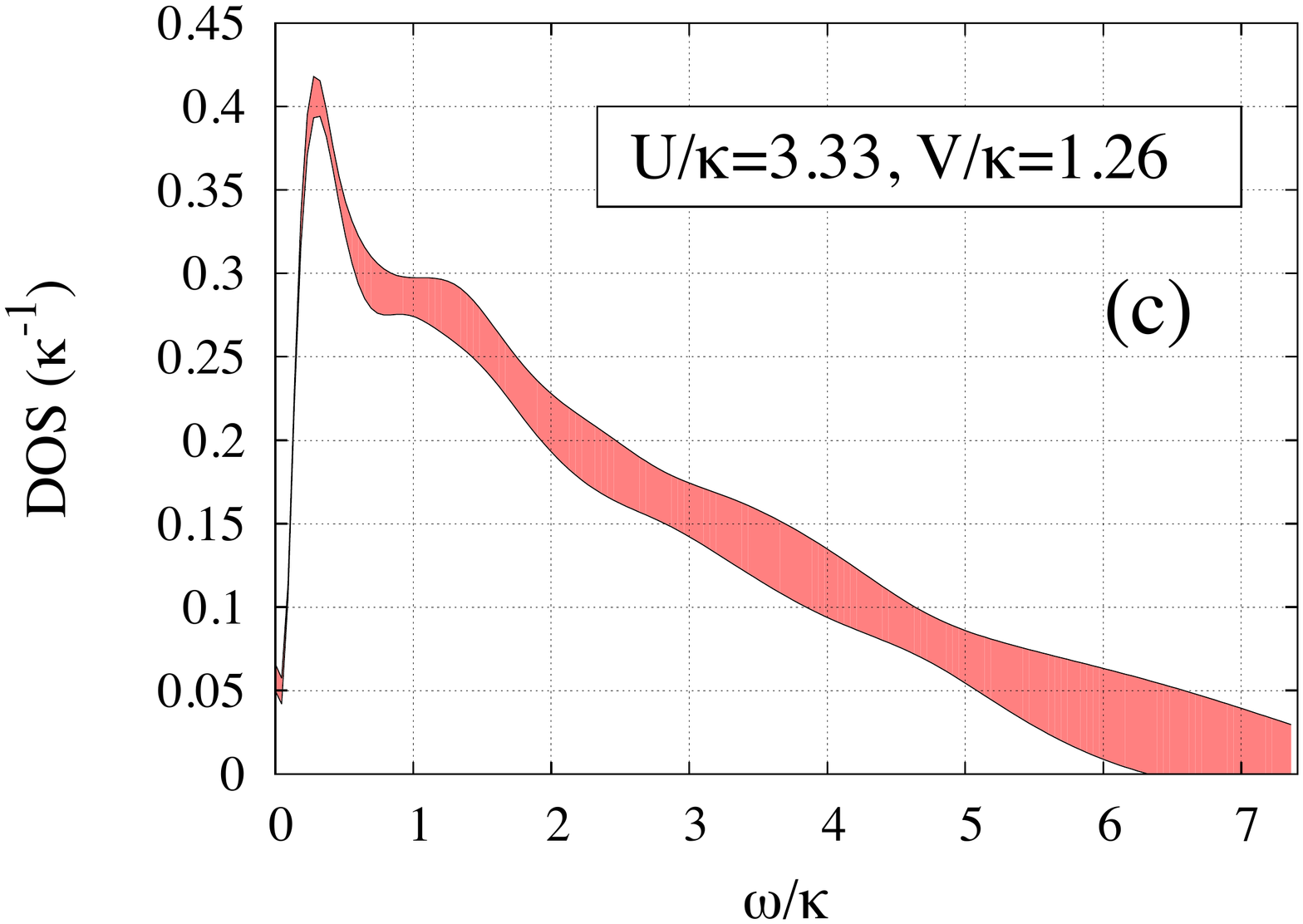}
        \caption{Plot of the DOS for the lattice ensemble with electron-electron interaction 
        $U/\kappa=0.83$, $V/\kappa = 0.5$ (a), $U/\kappa=1.66$, $V/\kappa = 0.62$ (b) and $U/\kappa=3.33$, $V/\kappa = 1.26$ (c). In all cases we have used a spatial lattice size of 
        $N_s=20$ with $N_{\tau}=160$ steps in Euclidean time and a temperature of $T/\kappa= 0.046$. Standard Tikhonov regularization with $\lambda=10^{-7}$  is employed during the analytical continuation. Thus, in all cases, we have the same resolution in frequency. No additional averaging in Euclidean time is applied.  Filled areas show the statistical error computed with the data binning procedure and the frequency $\omega$ corresponds to the center of the resolution function. The average value represents the estimator for DOS computed according to  (\ref{ConvolutionDeltaFunction}).     
        One can see the formation of the Hubbard bands, characterized by the peak at $E/\kappa \approx 0.5$, indicating that one is in the strongly-coupled phase for the largest interaction strength. For smaller interaction strengths we obtain a strong peak at zero energy indicating that the system is in the metallic phase.} 
        \label{fig:DOS1}
\end{figure}

This primary regularization is enough to obtain reasonably good results for DOS, as one can see from Fig.~\ref{fig:LambdaComparison}. Unfortunately, for the case of the charge density-density correlator, when we calculate the charge susceptibility by solving (\ref{GreenKuboDensity}) this regularization is not enough. The source of the this problem is the very large autocorrelation between different points in Euclidean time for the correlator $C({\bf q},\tau)$. As a result, it exhibits long-range fluctuations which can be seen in Fig.~\ref{fig:corr1}. We took the example correlator $C({\bf q},\tau)$ for the largest interaction strength considered ($U/\kappa=3.33$, $V/\kappa=1.26$) with the momentum $q$ directed along the $X-M$ line in the Brillouin zone. This is the most physically interesting case and will be discussed in detail in section \ref{sec:Results}.  After the application of the analytic continuation procedure, this oscillating correlator leads to a wildly fluctuating spectral function unless the regularization is so large that we can hardly resolve any structure due to very wide resolution functions $\delta(\omega_0,\omega)$. 

The way in which we have modified the BG method to alleviate this problem is through the introduction of ``interval averaging". In this procedure, we take the correlator data $\{ G(\tau_i), \Delta G(\tau_i); ~i=0,1,\dots , N_{\tau}-1 \}$ and map this to a new set $\{ \tilde{G}(\tilde{\tau}_j), \Delta \tilde{G}(\tilde{\tau}_j); ~j=1,\dots , N_{\text{int}} \}$ where
\beq
  \tilde{G}(\tilde{\tau}_j) &\equiv& \frac{1}{\tilde{N}_j} \sum^{\tilde{N}_j}_{i=1} G(\tau^{(j)}_i),\\ N_{\tau} &=& \sum^{N_{\text{int}}}_{j=1} \tilde{N}_j,~1 \leq \tilde{N}_j < N_{\tau}.  
\eeq 
 After performing the procedure, we are left with a set of averaged values of the correlator calculated over certain intervals in Euclidean time.  Due to the linearity of (\ref{GreenKuboDOS}) and (\ref{GreenKuboDensity}), one can construct $\{ \tilde{K}(\tilde{\tau}_j); ~j=1,\dots , N_{\text{int}} \}$ in an analogous manner and use these in the construction of the spectral function via the Tikhonov regularization. As a result, the spectral function will reproduce the averaged Euclidean correlator and will not follow the fluctuations within these intervals.

 \begin{figure}
        \centering
         \includegraphics[scale=0.3, angle=-90]{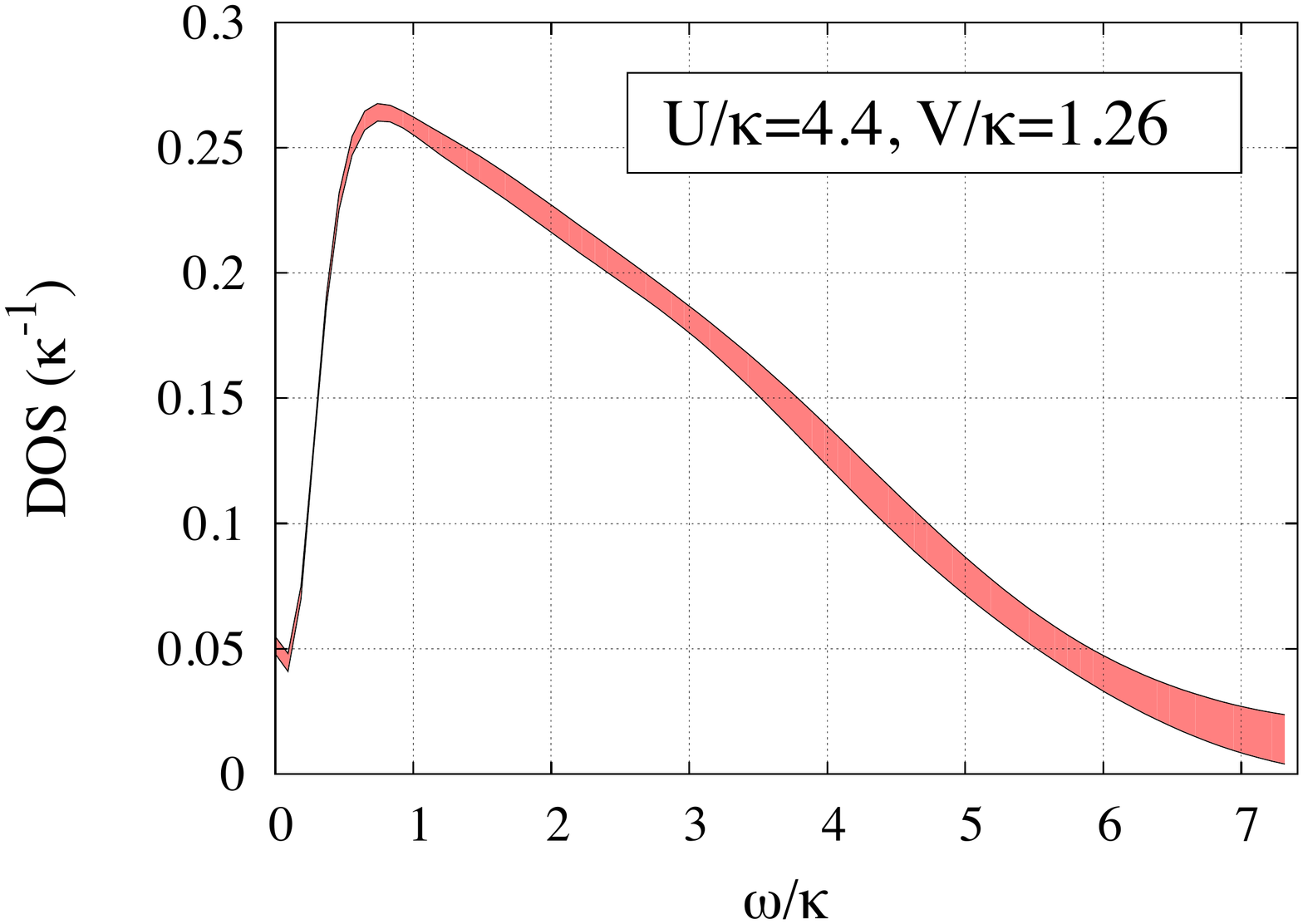}
        \caption{Plot of the DOS for the lattice ensemble with electron-electron interaction $U/\kappa=4.4$, $V/\kappa = 1.26$, which corresponds to the smallest interaction strength studied in \cite{vanLoon} ($U^{*}=1.1$ in their notation). We have used a spatial lattice size of 
        $N_s=20$ with $N_{\tau}=160$ steps in Euclidean time and two times larger temperature $T/\kappa= 0.092$, which is almost equal to the one used in \cite{vanLoon}. The set up for the analytic continuation is exactly the same as for Fig.~\ref{fig:DOS1}.
        One can see that in contrast to the results displayed in \cite{vanLoon}, the Hubbard bands are already formed, which suggests that the metal-insulator phase transition is shifted to a smaller interaction strength even at the same temperature.} 
        \label{fig:DOS_high}
\end{figure}

Typically, the signal-to-noise ratio of a Euclidean correlator $G(\tau)$ becomes worse as one approaches $\tau = \beta/2$. Furthermore, the correlations between adjacent points in Euclidean time are strong as one approaches this point. In light of this, we leave the points near $\tau=0$ untouched, while points closer to $\beta/2$ are bunched into intervals of longer length. 
In order to examine the dependence of the results on the choice of the intervals, we have performed the analytic continuation for the two different choices of intervals displayed in Fig.~\ref{fig:corr1}. The results of the analytic continuation are shown in Fig.~\ref{fig:compare_average}. One can see that the two agree within error bars which indicates that the qualitative features of the spectral function do not drastically change for different variants of averaging. In our calculations for the charge susceptibility we will use the first variant of averaging shown in Fig.~\textcolor{red}{4a}.

\begin{figure} 
        \centering
        \includegraphics[trim=1.5cm 3cm 0 0, scale=0.35, angle=0]{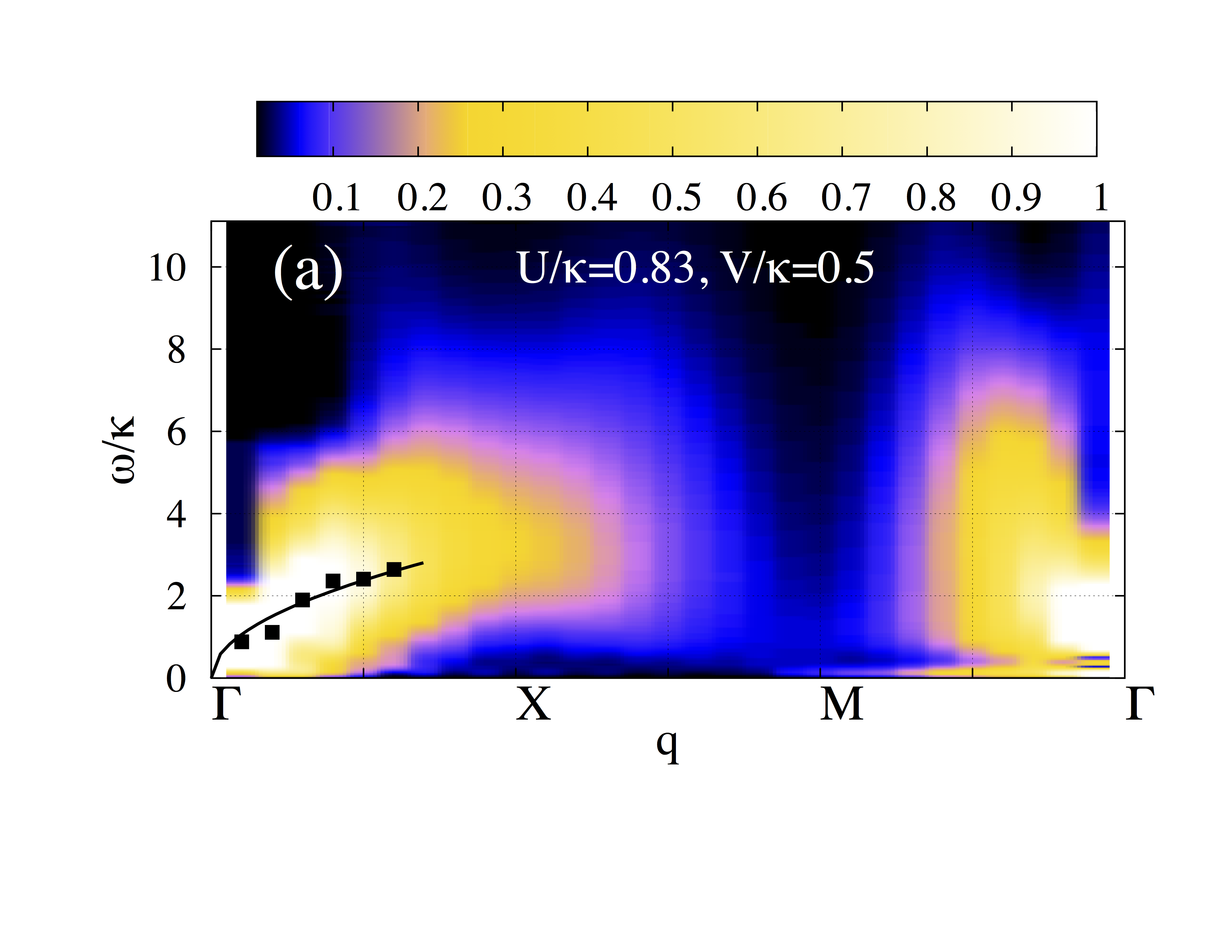}
        \includegraphics[trim=1.5cm 3cm 0 0, scale=0.35, angle=0]{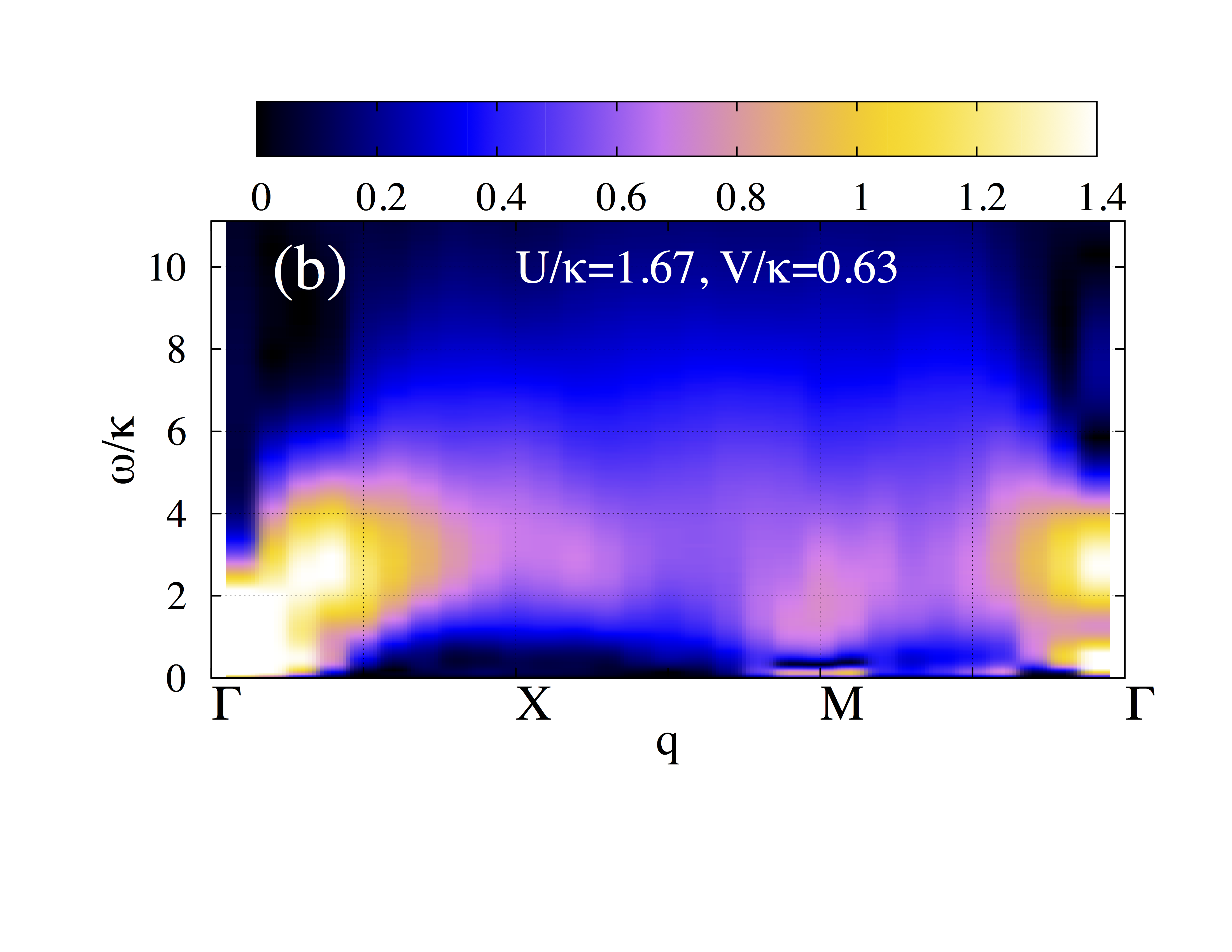}
        \includegraphics[trim=1.5cm 3cm 0 0, scale=0.35, angle=0]{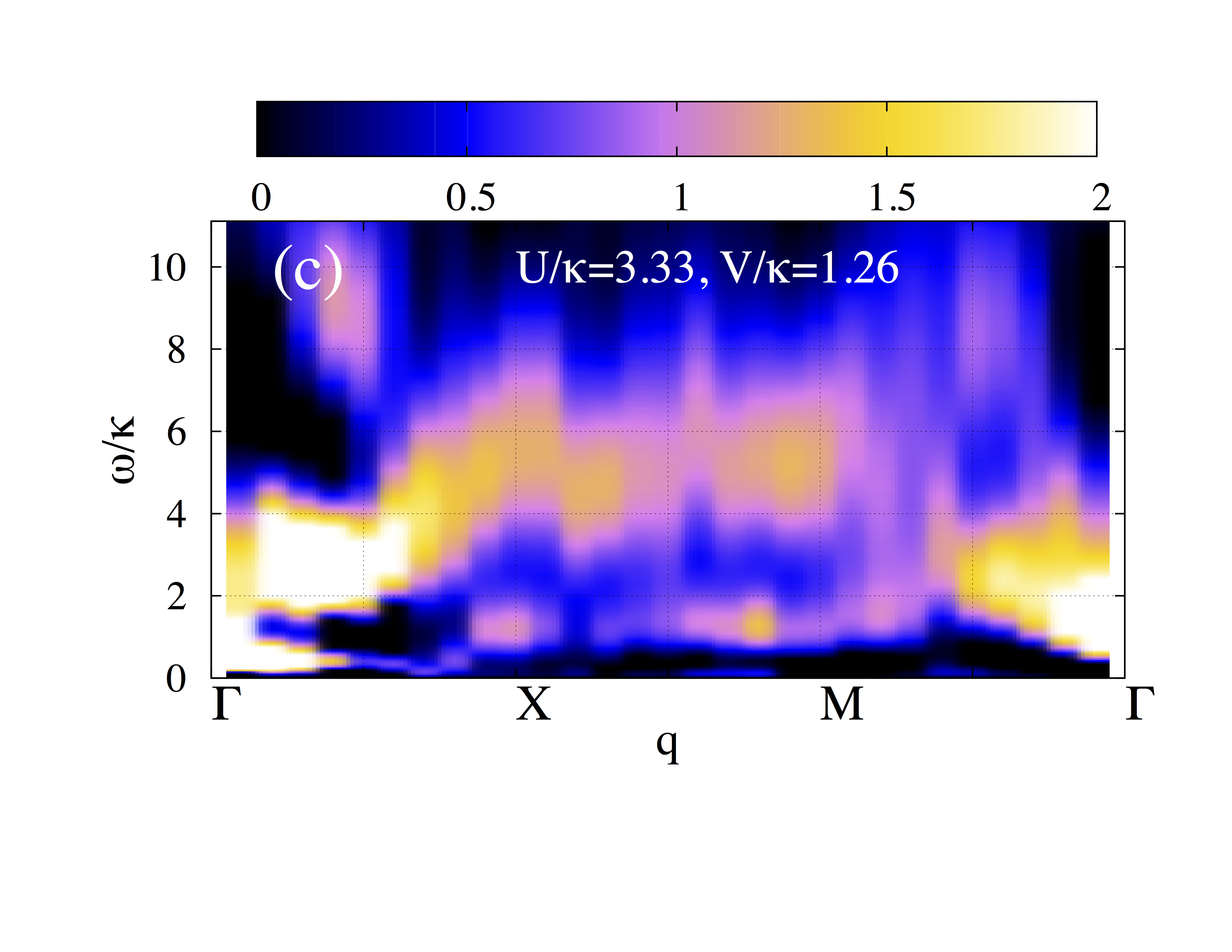}
        \caption{ Charge susceptibility for the square lattice Hubbard-Coulomb model for three different interaction strengths. The color scale corresponds to $V(q) \mbox{Im} \chi_\rho (\omega, q)$. For the spectral function  $\mbox{Im} \chi_\rho (\omega, q)$ we use the estimated value after analytic continuation (see (\ref{ConvolutionDeltaFunction})). The same lattice ensembles were used as in Fig.~\ref{fig:DOS1}. Standard Tikhonov regularization with $\lambda=5 \times 10^{-6}$ and additional time averaging according to Fig.~\textcolor{red}{4a} is used in all cases. 
       }
        \label{fig:Susceptibility1}
\end{figure}
 
 \section{\label{sec:Results}Results}
 In this section, we present our results for the spectral functions. 
 We perform all of our calculations with a spatial lattice size of $N_s=20$ and $N_\tau=160$ Euclidean time slices. Temperature is equal to $T=0.046$ in units of the hopping parameter $\kappa$. This temperature is smaller then the one used in \cite{vanLoon}, but we really need it since the resolution of the BG method is limited by temperature. In order to justify our conclusions we've made also one calculation for 2 times higher temperature $T=0.092 \kappa$. For each strength of the electron-electron interaction we have generated $\sim 10^3$ Hubbard field configurations for the calculation of the relevant observables. Since the width of the resolution functions (quantity $D$ in (\ref{ResolutionFunctionWidth})) is bounded from below by temperature, the spacing between neighboring values of $\omega_0$ is equal to the temperature in our calculations. The upper bound for $\omega_0$ is defined by the bandwidth, where all considered spectral functions go to zero.

 To make contact with the results of \cite{vanLoon}, we state the values of the parameters characterizing the interaction potential in our notation. The three sets of parameters used in \cite{vanLoon} correspond to $V \approx 1.26$ and $U = 4.4, 8.2, 10.4$ (in units of the hopping parameter $\kappa$). In \cite{vanLoon} the Mott transition was observed somewhere between $U = 8.2$ and $U=10.4$. Our first aim is to identify the real position of the phase transition from our non-perturbative Monte-Carlo calculations. For this purpose , we calculate the DOS for several pairs of $U$ and $V$ which characterize the electron-electron interaction. The results from these calculations are presented in Fig.~\ref{fig:DOS1}.
 
 One can clearly see that even for the case of $U = 3.33$, $V=1.26$ (see Fig. \textcolor{red}{6c}), which is smaller then the smallest interaction strength from \cite{vanLoon}, the system is already in the insulating state.  In this regime, the Hubbard bands have already formed and the quasiparticle weight at $\omega=0$ practically vanishes (due to finite temperature and the finite width of resolution functions it can not vanish completely). In Fig. \textcolor{red}{6b}, the DOS for the couplings $U = 1.66$, $V=0.62$ demonstrates that one is now firmly in the regime with well-defined quasiparticles at $\omega=0$. Finally, Fig.~\textcolor{red}{6a} shows the DOS for $U=0.83$, $V=0.5$ which is deep in the metallic phase. These results show that the EDMFT analysis from \cite{vanLoon} overestimates the critical coupling $U_c$ of the Mott transition for the Hubbard-Coulomb model.We emphasize the fact that this is not a temperature effect, as it is known that for the square-lattice Hubbard model, the phase transition is shifted to smaller $U$ as the temperature decreases \cite{FateMottHubbard}. The results displayed in Fig.~\ref{fig:DOS_high} show that, at twice the temperature of the ensembles in Fig.~\ref{fig:DOS1}, we are already in the insulating phase. For these same values of the coupling, $U = 4.4$, $V=1.26$, and a slightly smaller temperature, the authors of \cite{vanLoon} find the system to be in the metallic phase.

 Another difference is that one can not confirm the situation reported in \cite{vanLoon} for intermediate coupling: near the phase transition the DOS had equally high peaks at zero energy and at some non-zero values $\pm E_0$. It is possible that something similar develops in the Fig.~\textcolor{red}{6b}, but the second peak is too small and is basically within the error bars.

\begin{figure}
        \centering
        \includegraphics[scale=0.3, angle=270]{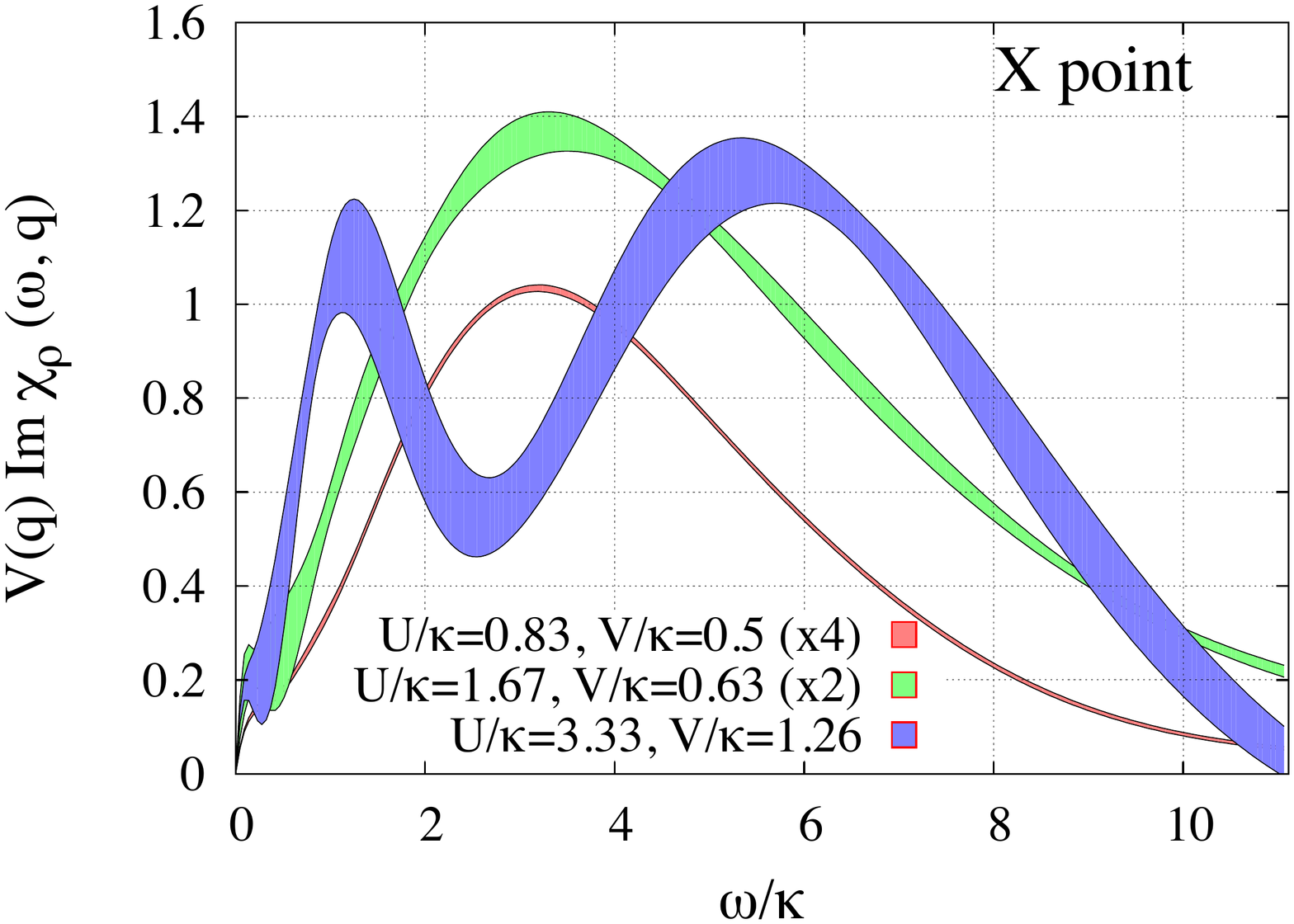}
              \caption{Comparison of the spectral function for the charge-charge correlator at the X point for different interaction strengths. The lattice set up and the parameters of the analytic continuation procedure are identical to the ones used for Fig.~\ref{fig:Susceptibility1}. The filled areas show the statistical error computed with the data binning procedure and the frequency $\omega$ corresponds to the center of the resolution function. The spectral functions for smaller interaction strengths are rescaled by factors of 2 and 4 in order to fit to the same scale.
              }
        \label{fig:compare_X}
\end{figure}

The results for the charge susceptibility in momentum space are presented in Figs.~\ref{fig:Susceptibility1} and \ref{fig:compare_X} for the same lattice set up and interaction strengths as was presented for DOS in Fig.~\ref{fig:DOS1}. The plots start from the center of the Brillouin zone ($\Gamma$-point), continues out towards the edge of the BZ along the $k_x$-axis, moves along the edge of the BZ parallel to the $k_y$-axis, and then finally returns to the center of the BZ along the diagonal connecting the point $M \equiv (\pi,\pi)$ with $\Gamma$.  For the smallest interaction strength we have tried to reproduce the $\sqrt{q}$ dispersion relation in the vicinity of the $\Gamma$-point (see Fig.~\textcolor{red}{8a}). The positions of the peaks at a given value of the momentum are depicted with black dots and we have fitted their position with the function  $f(q)=C \sqrt{q}$. Despite the fact that the statistical fluctuations in the Monte Carlo data do not allow us to fully justify this fitting procedure, our data at least do not contradict the $\sqrt{q}$ dispersion at low interaction strength. When we move to larger $U$ and $V$, the dispersion relation sufficiently changes and at the largest interaction strength, already in the insulating state, it finally splits into two branches. The splitting is most prominent along the $X-M$ line in the Brillouin zone. The case of the $X$-point is shown in the Fig.~\ref{fig:compare_X} separately for all three strengths of the electron-electron interaction. One can clearly see two peaks near $\omega=\kappa$ and $\omega=5\kappa$.

 \section{\label{sec:Conclusion}Conclusions}
We have studied the Mott metal-insulator transition in the Hubbard-Coulomb model on the square lattice using unbiased quantum Monte Carlo calculations on finite clusters. The hybrid Monte Carlo  algorithm was used to effectively study the system which contained a long-range interaction. In order to obtain the real-frequency spectral functions, we have developed a modified version of the Backus-Hilbert method for analytic continuation from Euclidean to real time. The metal-insulator phase transition was observed directly by the calculation the density of states. The decrease of DOS at zero energy and the formation of the Hubbard bands was observed across the phase transition. It was observed that the position of the phase transition is sufficiently shifted towards smaller interaction strength in comparison to previous EDMFT predictions. 

The behavior of the momentum-resolved charge susceptibility across the phase transition was also studied. This data was used to reveal the dispersion relation of the plasmons both in the metallic and in the insulating phase. The main aim was to check the predictions from \cite{vanLoon} concerning the splitting of the plasmonic dispersion relation into two bands in the region of the interaction strength close to the phase transition. The splitting was indeed observed in our Monte Carlo data. However, according to our calculations this phenomena tends to emerge in the situation when the material is already in the insulating state.

The data presented in the paper can be used as a benchmark in the further development of methods for strongly-correlated systems with long-range interactions. 

The modified BG method developed in this paper can also be used in cases where a non-biased estimate for the spectral function is important. The code used in the current paper is now accessible online \cite{BG_code1}.

\section*{Acknowledgements}
MU acknowledges inspiring and fruitful discussions
with Prof.~Mikhail Katsnelson. The work of MU
was supported by DFG grant BU 2626/2-1. SZ acknowledges support by the National Science Foundation (USA) under grant PHY-1516509 and by the Jefferson Science Associates,
LLC under U.S. DOE Contract $\#$ DE-AC05-06OR23177.
This work was partially supported by the HPC Center of Champagne-Ardenne ROMEO. CW acknowledges the warm hospitality of the University of Regensburg and would like to thank Pavel Buividovich for his financial support and for discussions in the early stages of the project. CW is supported by the University of Kent, School of Physical Sciences. 


\end{document}